%% file: resolution.tex
\documentclass{elsart}
\usepackage{graphicx}
\usepackage{amsmath}
\usepackage{amssymb}

\input{defs}

\begin{document}

\journal{Nucl. Inst. and Meth. A}

%\begin{flushright}
%\today
%\end{flushright}

\begin{frontmatter}

\title{Proper-time Resolution Function for
Measurement of  Time Evolution of $B$ Mesons  at the 
KEK  \textit{B}-Factory}
\author{H.~Tajima},
\author{H.~Aihara}\footnote{Corresponding author.\\ Email address: aihara@phys.s.u-tokyo.ac.jp},
\author{T.~Higuchi},
\author{H.~Kawai},
\author{T.~Nakadaira},
\author{J.~Tanaka}, 
\author{T.~Tomura}, and 
\author{M.~Yokoyama}
\address{Department of Physics, University of Tokyo \\ 7-3-1 Hongo, Bunkyo-ku, Tokyo, Japan}
\author{M.~Hazumi},
\author{Y.~Sakai}, and
\author{K.~Sumisawa}
\address{High Energy Accelerator Research Organization (KEK)\\
1-1 Oho, Tsukuba, Ibaraki, Japan}
\author{T.~Kawasaki}
\address{Graduate School of Science and Technology, 
Niigata University\\ 8050 Ikarashi Ni-no-cho, Niigata, Niigata, Japan}

\begin{abstract}
The proper-time resolution function for the measurement of the time evolution of $B$ mesons 
with the Belle detector at KEKB  
is studied in detail. 
The obtained resolution function is  applied to the measurement of $B$ meson lifetimes, 
the $B^0\overline{B}{}^0$ oscillation frequency and time-dependent \CP\ asymmetries.
\end{abstract}


\end{frontmatter}

\input{intro}

\input{belle}

\input{vertex}

\input{resfunc}

\input{application}

\input{conclusion}

\section*{Acknowledgment}
We would like to thank the members of  the Belle collaboration,
in particular, those of the SVD group for their effort to design, construct,
operate and maintain the SVD.
We are grateful to D.~Marlow  for careful reading of this article.
This work was supported in part by Grant-in-Aid for Scientific Research on
Priority Areas (Physics of CP violation) from Ministry of Education, Culture,
Sports, Science and Technology of Japan.

\input{biblio}

\end{document}

%% file: defs.tex
\newcommand*{\tbzresultmean}{1.554}
\newcommand*{\tbzresultstat}{\pm 0.030}
\newcommand*{\tbzresultsyst}{0.019}

\newcommand*{\tbmresultmean}{1.695}
\newcommand*{\tbmresultstat}{\pm 0.026}
\newcommand*{\tbmresultsyst}{0.015}

\newcommand*{\rbmresultmean}{1.091}
\newcommand*{\rbmresultstat}{\pm 0.023}
\newcommand*{\rbmresultsyst}{0.014}

\newcommand*{\tbzresult}{\tbzresultmean \tbzresultstat ({\rm stat}) \pm \tbzresultsyst ({\rm syst})}
\newcommand*{\tbmresult}{\tbmresultmean \tbmresultstat ({\rm stat}) \pm \tbmresultsyst ({\rm syst})}
\newcommand*{\rbmresult}{\rbmresultmean \rbmresultstat ({\rm stat}) \pm \rbmresultsyst ({\rm syst})}

\newcommand*{\degree}{{}^{\circ}}
\newcommand*{\micron}{\mu {\rm m}}
\newcommand*{\Gevc}{{\rm GeV}/c}

\newcommand*{\bzb}{{\overline{B}{}^0}}
\newcommand*{\bz}{{B^0}}
\newcommand*{\bm}{{B^-}}

\newcommand*{\bbar}{\overline{B}}

\newcommand*{\jpsi}{{J/\psi}}
\newcommand*{\UPS}{\Upsilon(4S)}
\newcommand*{\ks}{K_S^0}
\newcommand*{\kl}{K_L^0}
\newcommand*{\km}{K^-}
\newcommand*{\dz}{D^0}
\newcommand*{\dplus}{D^+}

\newcommand*{\pip}{\pi^+}
\newcommand*{\pim}{\pi^-}
\newcommand*{\rhom}{\rho^-}
\newcommand*{\dstarp}{D^{*+}}
\newcommand*{\kstarzb}{\overline{K}{}^{*0}}

\newcommand*{\bzbpsiks}{\bzb\to\jpsi\ks}
\newcommand*{\bmpsikm}{\bm\to\jpsi\km}
\newcommand*{\bdpi}{\bbar\to D\pi}

\newcommand*{\dt}{\Delta t}
\newcommand*{\delz}{\Delta z}
\newcommand*{\dtp}{\dt^\prime}
\newcommand*{\dtpp}{\dt^{\prime\prime}}
\newcommand*{\dtppp}{\dt^{\prime\prime\prime}}

\newcommand*{\Rnp}{R_{{\rm np}}}
\newcommand*{\Rk}{R_{{\rm k}}}
\newcommand*{\Rrec}{R_{{\rm ful}}}
\newcommand*{\fp}{f_\mathrm{p}}
\newcommand*{\Ep}{E_\mathrm{p}}
\newcommand*{\En}{E_\mathrm{n}}

\newcommand*{\taunp}{\tau_{\rm np}}
\newcommand*{\taunpp}{\taunp^\mathrm{p}}
\newcommand*{\taunpn}{\taunp^\mathrm{n}}
\newcommand*{\taup}{\tau_\mathrm{p}}
\newcommand*{\taun}{\tau_\mathrm{n}}

\newcommand*{\fsig}{f_{{\rm sig}}}

\newcommand*{\Psig}{P_{{\rm sig}}}
\newcommand*{\Pbg}{P_{{\rm bkg}}}
\newcommand*{\calPsig}{\mathcal{P}_{{\rm sig}}}

\newcommand*{\Rsig}{R_{{\rm sig}}}
\newcommand*{\Rbg}{R_{{\rm bkg}}}

\newcommand*{\Pol}{P_{{\rm ol}}}

\newcommand*{\fol}{f_{{\rm ol}}}

\newcommand*{\sigol}{\sigma_{{\rm ol}}}

\newcommand*{\Brec}{B_{\rm ful}}

\newcommand*{\trec}{t_{\rm ful}}
\newcommand*{\tasc}{t_{\rm asc}}
\newcommand*{\zrec}{z_{\rm ful}}
\newcommand*{\zasc}{z_{\rm asc}}
\newcommand*{\zi}{z_q}
\newcommand*{\znorm}{\zasc}
\newcommand*{\znonp}{\zasc^{\rm noNP}}
\newcommand*{\Rasc}{R_{{\rm asc}}}
\newcommand*{\Ri}{R_q}

\newcommand*{\Sasc}{\sigma^z_{\rm asc}}
\newcommand*{\Si}{\sigma_q}
\newcommand*{\srec}{s_{\rm ful}}
\newcommand*{\sasc}{s_{\rm asc}}
\newcommand*{\si}{s_q}

\newcommand*{\smain}{s_{\rm main}}
\newcommand*{\stail}{s_{\rm tail}}
\newcommand*{\ftail}{f_{\rm tail}}

\newcommand*{\smainbg}{s_{\rm main}^{{\rm bkg}}}
\newcommand*{\stailbg}{s_{\rm tail}^{{\rm bkg}}}
\newcommand*{\ftailbg}{f_{\rm tail}^{{\rm bkg}}}

\newcommand*{\tbg}{\tau_{{\rm bkg}}}
\newcommand*{\mudelbg}{\mu_\delta^{{\rm bkg}}}
\newcommand*{\mutaubg}{\mu_\tau^{{\rm bkg}}}
\newcommand*{\fdelbg}{f_\delta^{{\rm bkg}}}

\newcommand*{\taubz}{{\tau_{\bzb}}}
\newcommand*{\taubm}{{\tau_{\bm}}}
\newcommand*{\rbm}{{\taubm/\taubz}}

\newcommand*{\bzdstpm}{\bzb\to\dstarp\pim}
\newcommand*{\bzdstrhom}{\bzb\to\dstarp\rhom}
\newcommand*{\bzdppm}{\bzb\to\dplus\pim}
\newcommand*{\bmdzpm}{\bm\to\dz\pim}
\newcommand*{\bzbpsikst}{\bzb\to\jpsi\kstarzb}
\newcommand*{\CP}{\textit{CP}}
\newcommand*{\chisq}{\chi^2}
\newcommand*{\chisqndf}{\chisq/{\rm n.d.f.}}

\newcommand*{\bgu}{(\beta\gamma)_\Upsilon}
\newcommand*{\bgrec}{(\beta\gamma)_{\rm ful}}
\newcommand*{\bgasc}{(\beta\gamma)_{\rm asc}}

\newcommand*{\Eb}{E_B^{{\rm cms}}}
\newcommand*{\pb}{p_B^{{\rm cms}}}
\newcommand*{\thetab}{{\theta_B^{{\rm cms}}}}
\newcommand*{\cosb}{\cos\thetab}

\newcommand*{\fcp}{f_{CP}}
\newcommand*{\ftag}{f_{\rm tag}}
\newcommand*{\ttag}{t_{\rm tag}}
\newcommand*{\Dt}{\Delta t}
\newcommand*{\dM}{\Delta m_d}
\newcommand*{\tcp}{t_{CP}}
\newcommand*{\sinbb}{\sin2\phi_1}

%% file: intro.tex
\section{Introduction}

KEKB is an asymmetric
electron-positron collider designed to produce boosted $B$ mesons~\cite{KEKB}.
At KEKB, electrons (8.0~GeV) and positrons (3.5~GeV) collide 
at a small $(\pm 11~{\rm mrad})$ crossing angle. 
Their annihilations produce $\Upsilon(4S)$ mesons 
moving nearly along the $z$ axis, defined as anti-parallel to the positron beam direction,
with a Lorentz boost factor of $(\beta\gamma)_\Upsilon=0.425$, 
and decaying  to $B^0\overline{B}{}^0$ or $B^+B^-$.
Precise determination of the proper-time interval $(\Delta t)$ between the 
two $B$ meson decays is essential 
for the measurement of $B$ meson lifetimes, the $\bz\bzb$ oscillation frequency, 
and time-dependent \CP\ asymmetries.
Since the $B$ mesons are nearly at rest in the $\UPS$ center of mass system
(cms), $\Delta t$ can be determined from the separation in $z$ 
between the two $B$ decay vertices ($\Delta z$).
The average $\Delta z$  at KEKB  is $c\tau_B(\beta\gamma)_\Upsilon \sim 200~\mu$m,
where $\tau_B$ is the $B$ meson lifetime.
In this article, we consider the analysis in which one of the $B$ decay vertices 
is determined from a fully reconstructed $B$ meson, 
while the other is determined
using the rest of the tracks in the event.

In order to extract the {\it true}  $\Delta t$   distribution from the observed $\Delta z$ distribution
it is necessary to unfold the vertex detector resolution and (possible) bias  in the measurement of $\delz$.
Because the detector resolution is of the same order as the average $\delz$
at KEKB, an understanding of the 
resolution is crucial for precise measurements.
For  measurement of time-dependent quantities, 
we employ an unbinned maximum likelihood method.
A probability density function for the likelihood fit is obtained as a convolution of
a theoretical $\dt$ distribution with the resolution function.

This paper describes the resolution function developed for the precise measurement of $B$ meson lifetimes with fully-reconstructed hadronic decay final states 
at the Belle experiment~\cite{blife}.
The obtained resolution function is also applied to the measurements of the $\bz\bzb$ oscillation frequency~\cite{mixing} 
and time-dependent \CP\ asymmetries~\cite{CPV}.

The organization of the paper is as follows: 
We give a brief description of the Belle detector in Section~\ref{sec:detector}.
In Section~\ref{sec:vertex}, the method of vertex reconstruction is described.
Details of the resolution function and its application to the data are described
in Sections~\ref{sec:resfunc} and \ref{sec:application},  respectively, followed by
a conclusion  in Section~\ref{sec:conclusion}.

%% file: belle.tex
\section{The detector}
\label{sec:detector}
\begin{figure}
\begin{center}
 \resizebox{11cm}{!}{\includegraphics*{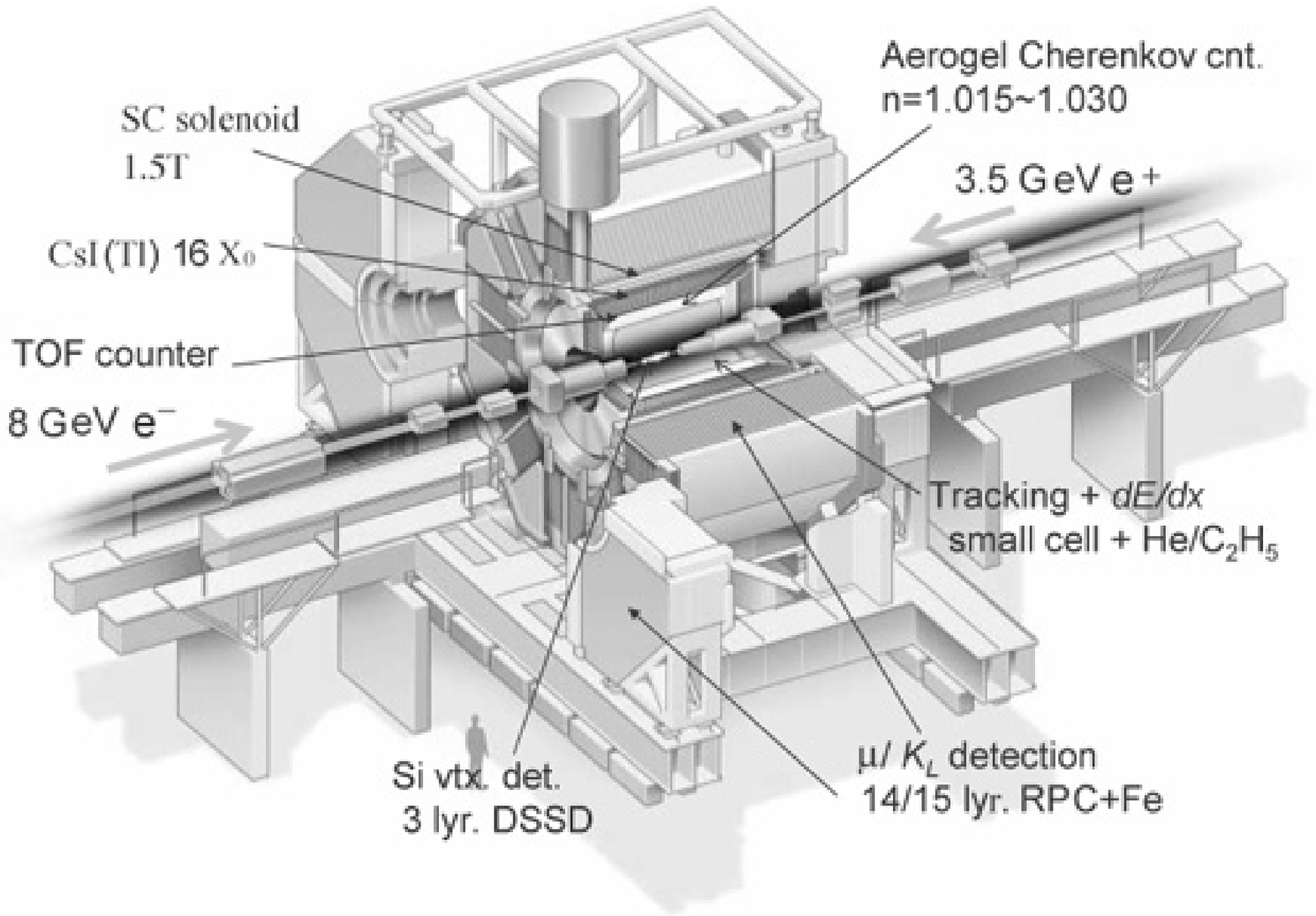}}
  \caption{\label{fig:belle} Belle detector.}
\end{center}
\end{figure}

The Belle detector, shown schematically  in Fig.~\ref{fig:belle}, is a large-solid-angle magnetic spectrometer that consists of 
a three-layer silicon vertex detector (SVD), 
a 50-layer central drift chamber, 
a mosaic of aerogel threshold  Cherenkov counters, 
time-of-flight scintillation counters, 
and an array of CsI(Tl) crystals located inside 
a superconducting solenoid  that provides a 1.5 T magnetic field. 
An iron flux return located outside of the coil is instrumented to detect $\kl$ mesons and to identify muons. 
The detector is described in detail elsewhere~\cite{belle-nim}.
 
The SVD, which plays an essential role in reconstructing decay vertices,  
consists of three concentric cylindrical layers of double-sided silicon strip detectors.
It covers the angular range of $23\degree < \theta < 139\degree$, where $\theta$ is the polar angle
from the $z$ axis.
This coverage corresponds to 86\% of the full solid angle in the cms.
The three layers are located at the radii of 3.0, 4.5, and 6.0~cm.
The strip pitches are 84~$\micron$ for the measurement of the $z$ coordinate and 25~$\micron$ 
for the measurement of azimuthal angle $\phi$.
The impact parameter resolutions for charged tracks~\cite{SVD_per} are measured to be 
$\sigma_{xy} = \sqrt{(19)^2 + (50/(p\beta\sin^{3/2}\theta))^2}\  \micron$ 
in the plane perpendicular to the $z$ axis and 
$\sigma_{z} =\sqrt{ (36)^2 + (42/(p\beta\sin^{5/2}\theta))^2}\  \micron$ 
along the $z$ axis, where $\beta = pc/E$, $p$ and $E$ are 
the momentum ($\Gevc$) and energy (GeV) of the particle.

%% file: vertex.tex
\section{Proper-time interval reconstruction}
\label{sec:vertex}
\begin{figure}
\begin{center}
 \resizebox{10cm}{!}{\includegraphics*{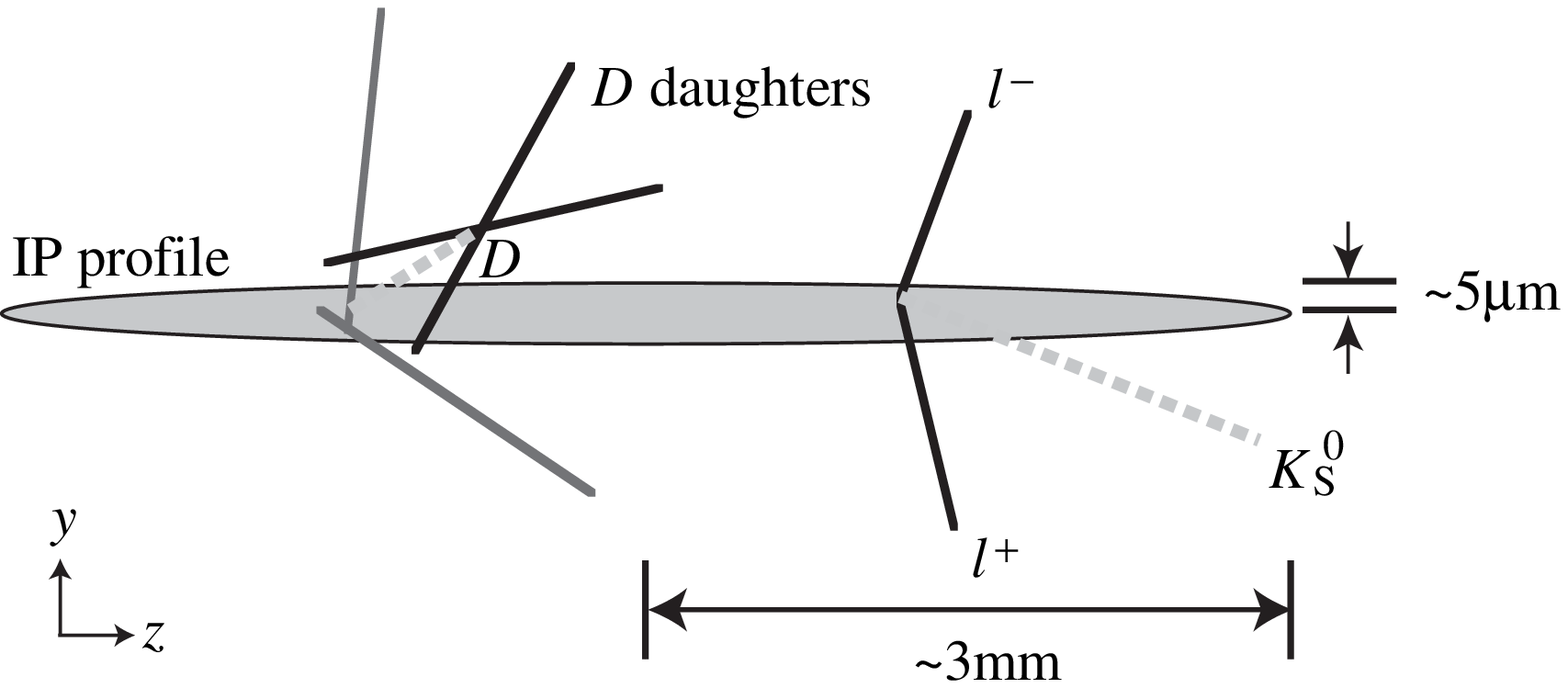}}
 \caption{\label{fig:vertex_recon}Illustration of vertex reconstruction
    of two $B$ decay vertices.}
\end{center}
\end{figure}

In this section, we describe the reconstruction  of the proper-time interval between
the decay points of 
the two $B$ mesons produced at KEKB.
Figure \ref{fig:vertex_recon} illustrates reconstruction of the decay vertices.
The decay vertices of the two $B$ mesons in each event are fitted
using tracks that 
have at least one three-dimensional coordinate determined from associated $r$-$\phi$ and $z$ hits 
in the same SVD layer plus one or more additional $z$ hits in other SVD layers.
%are associated with both $z$ and $r$-$\phi$ hits in at least one SVD
%layer and a $z$ hit in at least one additional layer.
We impose the constraint that they are consistent with
the interaction point (IP) profile, smeared in the $r$-$\phi$ plane
by $21~\micron$ to account for the transverse $B$ decay length.
The IP profile is described as a three-dimensional Gaussian,
the parameters of which are determined in each run
(in finer subdivisions, every 10,000 - 60, 000 events, for  the mean position)
using hadronic events.
The size of the IP region is typically $\sigma_x \simeq 100~\micron$,
$\sigma_y \simeq 5~\micron$, and $\sigma_z \simeq 3$~mm,
where $x$ and $y$ denote the horizontal and vertical directions, respectively.

One $B$ meson is fully reconstructed in one of the following decay 
modes~\footnote{Throughout this paper, when a decay mode is 
quoted the inclusion of charge conjugate mode is implied.}: 
$\bzb \to \dplus\pim$, $\dstarp\pim$, $\dstarp\rhom$, $\jpsi\ks$, $\jpsi\kstarzb$, 
$\bm \to \dz\pim$, and $\jpsi\km$, where
$\jpsi$ is reconstructed via  $\jpsi\to \ell^+\ell^- (\ell=e,\mu)$ decay,
and $D^{*+}$ via $D^{*+}\to D^0\pi^+$ decay.
Neutral and charged $D$ mesons are reconstructed in the following channels:
$D^0\to K^-\pi^+, K^-\pi^+\pi^0, K^-\pi^+\pi^+\pi^-$, and 
$D^+\to K^-\pi^+\pi^+$.
The details of the event selection can be found elsewhere~\cite{blife}.

In the case of a fully reconstructed $\bbar \to D^{(*)} X$ decay, 
the $B$ decay point is obtained from the vertex position and momentum vector of 
the reconstructed $D$ meson and tracks other than the slow $\pip$ candidate from $\dstarp$ decay.
For a fully reconstructed $\bbar \to \jpsi X$ decay, the $B$ vertex is determined
using lepton tracks from the $\jpsi$.

The decay vertex of the associated $B$ meson is determined by applying a vertex-fit program to all tracks 
not assigned to the fully reconstructed $B$ meson;
however, poorly reconstructed tracks 
(with a longitudinal position error in excess of $500~\micron$) as well as tracks that are 
likely to 
come from $\ks$ decays (forming the $\ks$ mass with another track, 
or emanating from a point more than $500~\micron$ 
away from the fully reconstructed $B$ vertex 
in the $r$-$\phi$ plane) are not used.
If the reduced $\chisq$ $(\chisqndf)$ associated with 
a found vertex exceeds 20, the track making the 
largest $\chisq$ contribution is removed and the vertex is refitted.
This procedure is repeated until $\chisqndf<20$ is obtained or only one track is left.
If, however,  the track to be removed is a lepton with a cms momentum 
greater than 1.1~$\Gevc$,  we keep the lepton and remove the track making
the second largest $\chisq$ contribution.
This  is 
because high-momentum leptons are 
likely to come from primary semi-leptonic $B$ decays.
The presence of a secondary charm $(b\to c)$ decay vertex
in the associated $B$ meson results in
a shift of the reconstructed vertex point 
toward the charm flight direction and degrades the vertex resolution.
A Monte Carlo (MC) simulation  study shows that
the shift and the resolution of the associated $B$ decay vertex 
are $\sim 20$~$\micron$ and $\sim 140$~$\micron$ (rms), respectively,
while the resolution of the fully reconstructed $B$ decay vertex is
$\sim 75$~$\micron$ (rms).

\begin{figure}
\begin{center}
 \resizebox{\textwidth}{!}{\includegraphics*{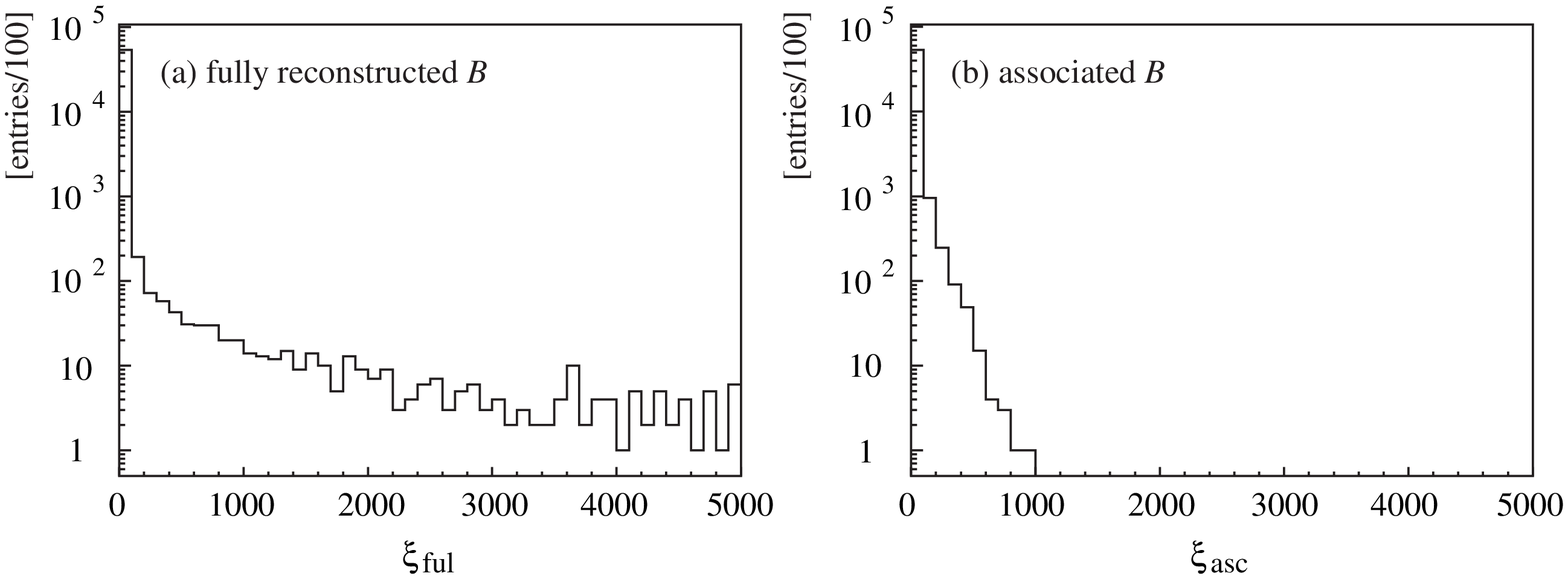}}
 \caption{\label{fig:xidist}The $\xi$ distributions for the (a) fully reconstructed 
and (b)  associated $B$ decays, obtained using $\overline{B}{}^0\to J/\psi \ks$ MC events.
%Because the  $\chisqndf$ requirement is imposed  on the associated $B$ decay vertices,
%the $\xi$ distribution for those vertices does not show an extended tail
%compared to that for the fully reconstructed $B$ decay vertices.
}
\end{center}
\end{figure}

\begin{figure}
\begin{center}
 \resizebox{\textwidth}{!}{\includegraphics*{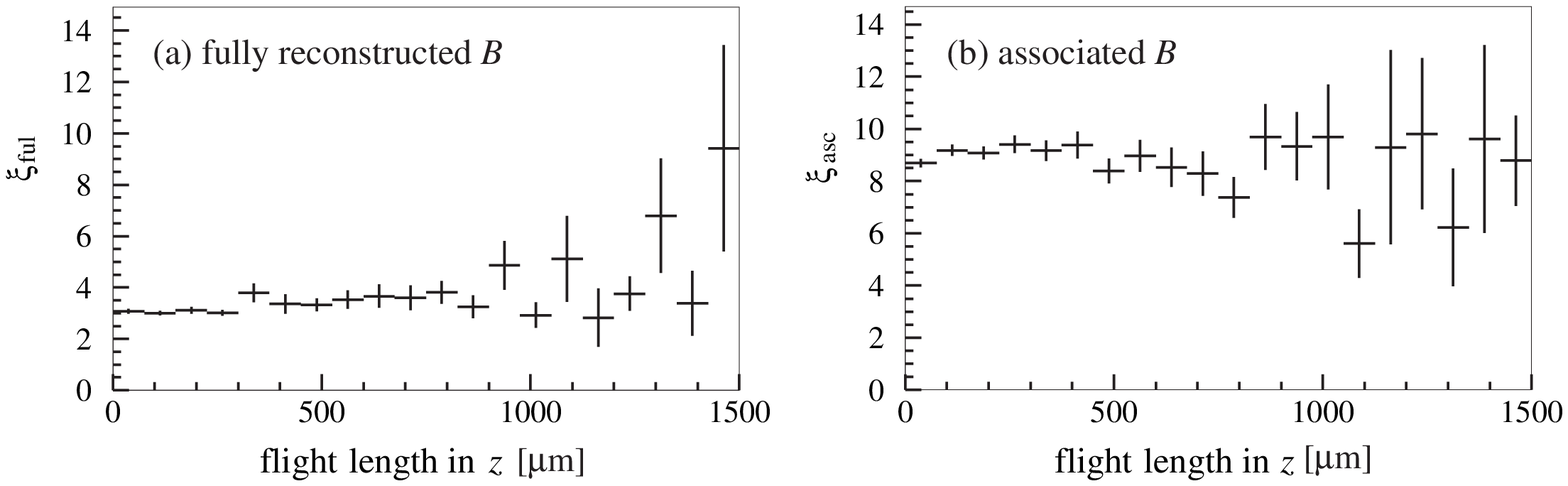}}
 \caption{\label{fig:xidecay}Monte Carlo $\xi$ distribution as a function of the $B$ decay length, 
obtained for the (a)  fully reconstructed  and (b) associated $B$ decay vertices. }
\end{center}
\end{figure}

Because of the IP profile constraint it is possible to reconstruct a decay vertex
even with a single track. 
The fraction of single-track vertices is $\sim 10\%$ and $\sim 22\%$ for the fully
reconstructed and associated $B$ decays, respectively.
For the multiple-track vertices, 
the quality of the vertex fit, for both the fully reconstructed and
associated $B$ meson decays, is further evaluated.
Using MC simulation,  we find that the vertex-fit $\chisq$ is correlated
with the $B$ decay length due to the tight IP constraint in the transverse plane.
To avoid this correlation, we use  the variable based on the $z$ information only 
(contrary to the vertex-fit $\chi^2$ calculated using three dimensional information):
\begin{equation}
\label{eq:xi}
\xi \equiv (1/2n) \sum^n_i \left[ (z_{\rm after}^i-z_{\rm before}^i) /
  \varepsilon_{\rm before}^i \right]^2,
\end{equation}
where $n$ is the number of tracks used in the fit
\footnote{For single-track vertices  $\xi$ cannot be defined.}, 
$z_{\rm before}^i$ and $z_{\rm after}^i$ are 
the $z$ positions of each track (at the closest approach to the origin)
before and after the vertex fit, respectively, and $\varepsilon_{\rm before}^i$ is the error of $z_{\rm before}^i$.
Figure~\ref{fig:xidist} shows the $\xi$ distributions for the (a) fully reconstructed and (b) associated $B$
decay vertices, obtained using a MC simulation.
Because the  $\chisqndf$ requirement is imposed  on the associated $B$ decay vertices,
the $\xi$ distribution for those vertices does not show an extended tail
compared to that for the fully reconstructed $B$ decay vertices.
Figure~\ref{fig:xidecay} shows the $\xi$ distributions as a function of the $B$ decay length for the (a) fully reconstructed
and (b) associated $B$ decay vertices.
As indicated, $\xi$ does not depend on the $B$ decay length. 
We require $\xi < 100$ for both vertices to eliminate poorly reconstructed vertices.
We find that
about 3\% of the fully reconstructed and 1\% of the associated $B$ decay vertices are rejected in the data.

The proper-time interval between the fully-reconstructed
and the associated $B$ decays is calculated
as 
\begin{equation}
\label{eq;deltat}
\dt =(\zrec - \zasc)/[c\bgu], 
\end{equation}
where $\zrec$ and $\zasc$
are the $z$ coordinates of the fully-reconstructed and associated
$B$ decay vertices, respectively.
We reject a small fraction ($\sim$0.2\%) of the events by requiring
$|\dt| < 70$~ps ($\sim$45$\tau_B$).

%% file: resfunc.tex
\section{Resolution function}
\label{sec:resfunc}

\subsection{Overview}
We extract the lifetimes of $B$ mesons using an unbinned maximum likelihood fit
to the observed $\dt$ distributions.
We maximize the likelihood function $L = \prod_i P(\dt_i)$,
where $P(\dt_i)$ is the probability density function (PDF) for
the  $\dt$ of  the $i$-th event and
the product is taken over all the selected events.

The function $P(\dt)$, expressed  as
\begin{equation}
P(\dt)=(1-\fol)\left[\fsig\Psig(\dt)+(1-\fsig)\Pbg(\dt)\right]+\fol\Pol(\dt),
\label{eq:pdf}
\end{equation}
contains contributions from the signal and the background ($\Psig$ and $\Pbg$),
where $\fsig$ is the signal purity determined on an event-by-event basis,
$\Psig$ is described as the convolution of a true PDF ($\calPsig$) with a resolution function ($\Rsig$)
and $\Pbg$ is expressed in a similar way:
\begin{equation}
P_{\rm sig (bkg)}(\dt)=\int_{-\infty}^{+\infty}\d(\dtp){\mathcal P}_{\rm sig(bkg)}(\dtp)R_{\rm sig(bkg)}
(\dt-\dtp).
\end{equation}
To account for a small number of events that give large $\dt$ in both the signal and background
(outlier components), we introduce a fraction of outliers ($\fol$)  and  a Gaussian function ($\Pol(\dt)$) to model
its distribution.
The true PDF for the signal, $\calPsig(\dt; \tau_B)$, is given by
\begin{equation}
  \label{eq:signal}
  \calPsig(\dt; \tau_B) = \frac{1}{2\tau_B}
  \exp\left(-\frac{|\dt|}{\tau_B}\right) ,
\end{equation}
where $\tau_B$ is, depending on the reconstructed mode in the event, 
either the  $\overline{B}{}^0$ or the $B^-$ lifetime.
The signal PDFs for the measurements of the $B^0\overline{B}{}^0$
oscillation frequency and the time-dependent $CP$ asymmetry
are given in Section 5.

The resolution function of the signal is constructed as the convolution of four different
contributions: the detector resolution on $\zrec$ and $\zasc$ ($\Rrec$ and $\Rasc$),
an additional smearing on $\zasc$ due to the inclusion of tracks which do not
originate from the associated $B$ vertex ($\Rnp$), mostly due to charm and $\ks$ decays,
and 
the kinematic approximation that the $B$ mesons are at rest
in the cms ($\Rk$).
The overall resolution function, $\Rsig(\dt)$, is expressed as
\begin{eqnarray}
  \Rsig (\dt)& = &
  \int \!\!\! \int \!\!\! \int_{-\infty}^{+\infty} \mathrm{d}(\dtp) \, \mathrm{d}(\dtpp) \,
\mathrm{d}(\dtppp) \, 
  \Rrec (\dt-\dtp)\Rasc(\dtp-\dtpp) \nonumber \\
 & & \times \Rnp(\dtpp-\dtppp) \Rk(\dtppp).
\end{eqnarray}

We use a MC simulation to understand the resolution function and determine its functional form.
The MC events are generated using the QQ event generator~\cite{qq} and the response of the Belle detector is modeled by a GEANT3-based full-simulation program~\cite{geant}.
One of $B$ mesons in each event is forced to decay to $\jpsi\ks$,  $\jpsi\km$ or $D\pi$ 
while the other decays generically to one of all possible final states.

\subsection{Detector resolution}
\begin{figure}
  \resizebox{\textwidth}{!}{\includegraphics{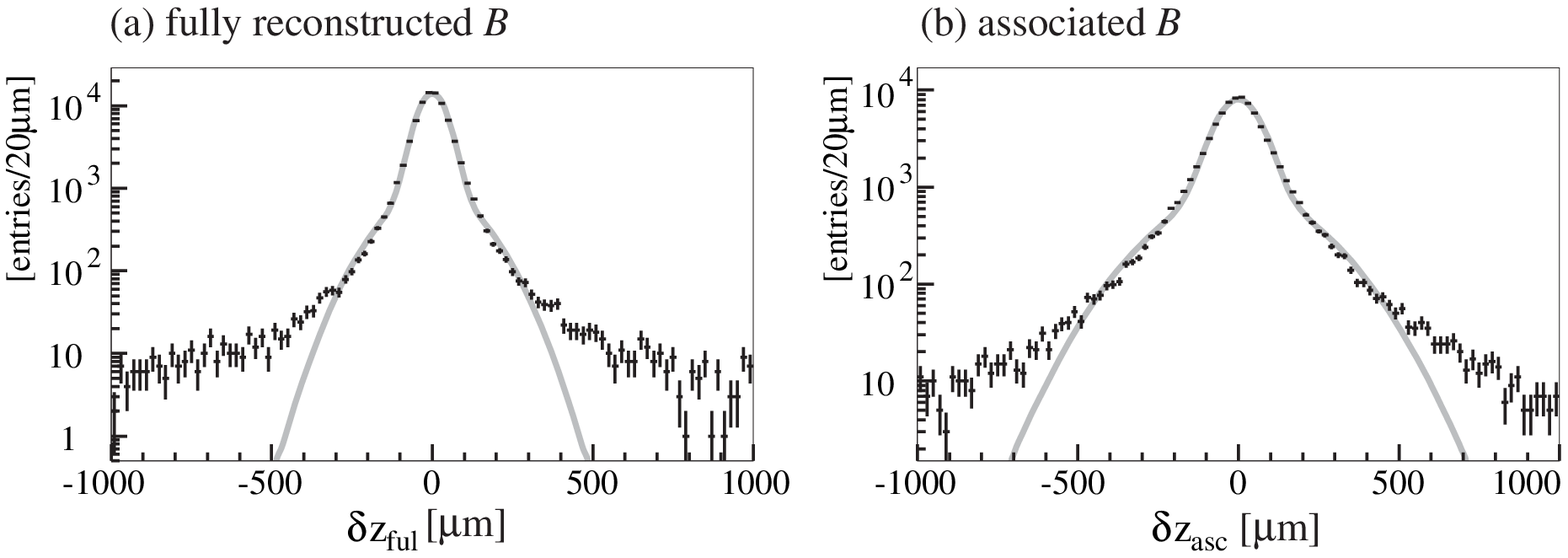}}
  \caption{\label{fig:poor} The $\delta z$ distributions of  the (a) fully reconstructed
    and (b) associated $B$ vertices obtained from a $\bzbpsiks$ MC sample.
    Superimposed are the results of a fit to the sum of  two Gaussians.  }
\end{figure}

The detector resolution ($\Rrec$ and $\Rasc$) is studied 
using a special MC simulation in which all short-lived ($\tau<10^{-9}$~s) secondary particles (including $\ks$ and $\Lambda$)  are forced to decay with zero lifetime
at the $B$ meson decay points.
Figures~\ref{fig:poor} (a) and (b) show the distributions of the difference in $z$ between the reconstructed
and generated vertex positions:
\begin{equation}
\delta \zi = \zi^{\rm rec} - \zi^{\rm gen} ,
\end{equation}
where $q= {\rm ful} ({\rm asc})$ is for the fully reconstructed (associated) $B$ vertex,
and the superscripts `rec' and `gen' denote the reconstructed and generated vertex positions, respectively.
Results of the fit  to a sum of two Gaussians are also shown.
The fitted curves do not represent the $\delta z$ distributions in the tail regions.
We also find that even a sum of three or more Gaussians with {\it constant} standard deviations 
cannot represent $\delta z$ properly. 
We therefore consider a more elaborate function that uses the {\it vertex-by-vertex} 
$z$-coordinate error of the
reconstructed vertex, $\sigma^\zi$ ($q$ = full, asc), 
as an input parameter.
The value of $\sigma^\zi$ is computed from  the error matrix of the tracks  used in the vertex fit
and the size of the IP region.
To construct  functional forms of $\Rrec$ and $\Rasc$
we investigate the distribution of  a pull,  defined as $\delta z_q$ divided by $\sigma^\zi$.
If the $\sigma^\zi$ estimation is correct on average,
the pull distribution is expected to be 
a single Gaussian with the standard deviation of unity.

\begin{figure}
  \resizebox{\textwidth}{!}{\includegraphics{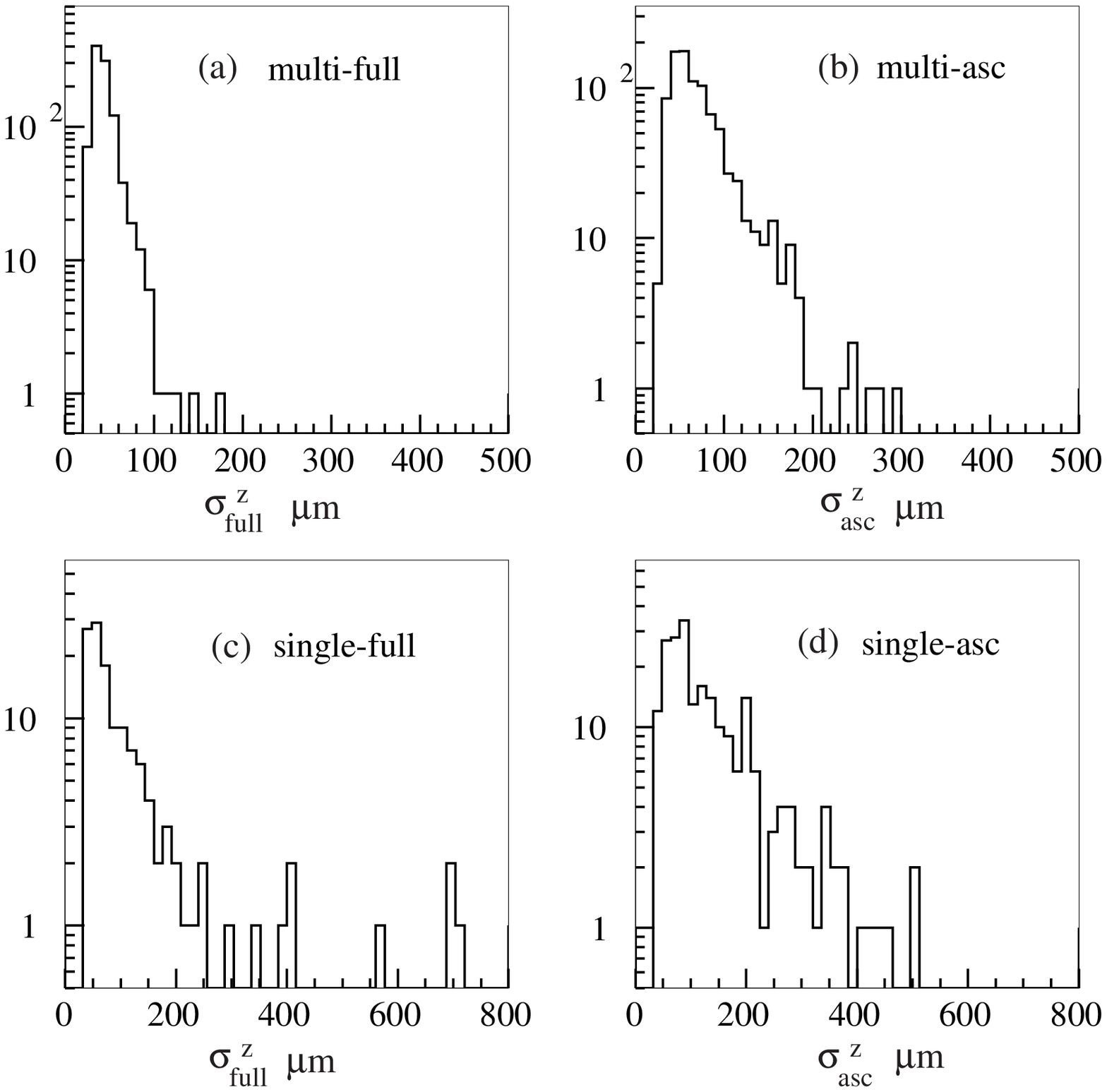}}
  \caption{\label{fig:evebyeve} Distributions of the vertex-by-vertex $z$-coordinate error 
$\sigma^z$ of the (a) (b) mutli-track and (c) (d) single-track vertices for  the fully reconstructed
    and associated $B$ decays, 
    obtained from $\overline{B^0}\to J/\psi K_S^0$ data.
}
\end{figure}

Because the resolution for the multiple-track vertices is better than that for the single-track vertices,
we consider them separately.
Figure~\ref{fig:evebyeve} shows the distributions of  $\sigma^z_{\rm full}$ and $\sigma^z_{\rm asc}$ for the multi-track
and single-track vertices obtained from $B^0\to J/\psi K_S^0$ data.


\subsubsection{Multiple-track vertex}
\begin{figure}
  \resizebox{\textwidth}{!}{\includegraphics{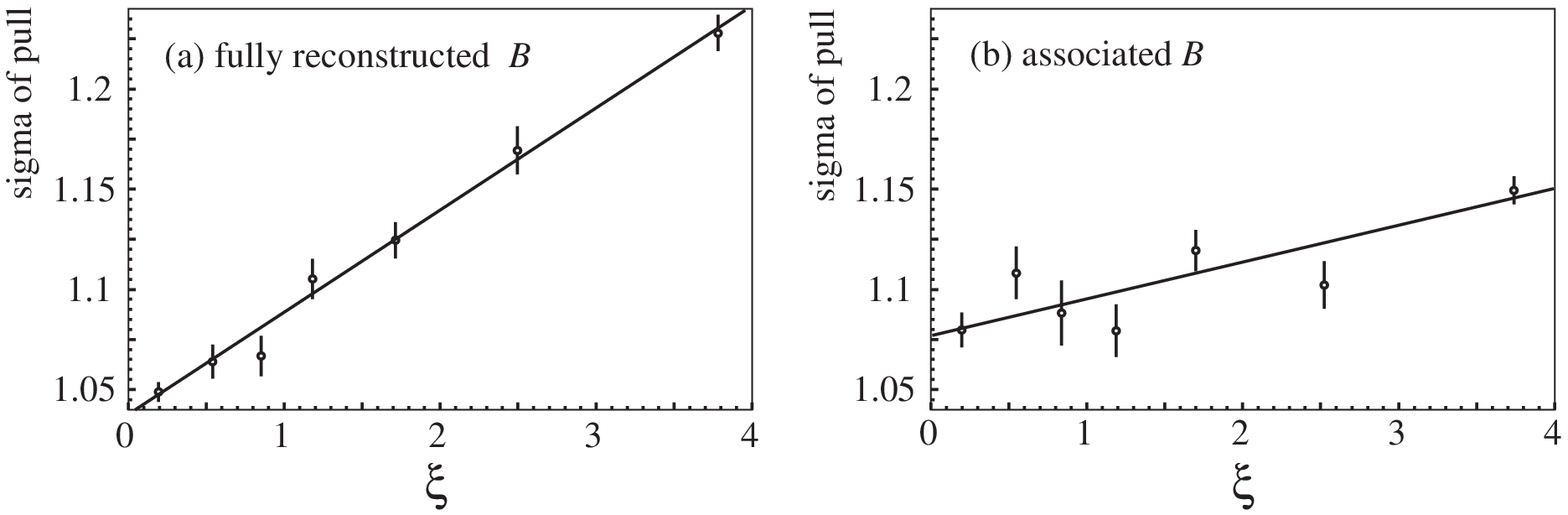}}
  \caption{\label{fig:lindep_s} Standard deviations of the pull
    distributions as a function of $\xi$ for (a)  the fully reconstructed 
    and (b) the associated  $B$ meson vertices.
    The distributions are obtained from a $\bzbpsiks$ MC sample.}
\end{figure}

We investigate the vertex fit quality dependence of the resolution
using the value of $\xi$ defined in Eq.~(\ref{eq:xi}).
We find that a pull distribution for vertices with similar $\xi$ values can be expressed as a single Gaussian.
%This can be seen  in Fig.~\ref{fig:singleG_xi}
%which shows the pull distributions
%for seven  different $\xi$ ranges.
%Results of a fit to a single Gaussian for each $\xi$ range are superimposed.
Furthermore, we find that the standard deviation of the distribution has a linear dependence on $\xi$
as shown in Fig.~\ref{fig:lindep_s}.
%  \begin{figure}
%    (a) \\
%  \begin{center}
%    \resizebox{0.9\textwidth}{!}{\includegraphics{fig/fit_to_singleG-rec.eps}} \\
%  \end{center}
%    \vspace*{5mm} 
%    (b) \\
%  \begin{center}
%    \resizebox{0.9\textwidth}{!}{\includegraphics{fig/fit_to_singleG-asc.eps}}
%  \end{center}
%    \caption{\label{fig:singleG_xi} Pull ($\delta z /\sigma^z$) distributions of (a) fully reconstructed and 
%    (b) associated $B$ vertex, for each $\xi$ range. 
%    Only vertices reconstructed with multiple tracks are shown.
%    Results of a fit to a single Gaussian are superimposed. 
%    These distributions are obtained from a $\bzbpsiks$ MC sample.}
%  \end{figure}

Results from this MC study lead 
us to model the detector resolution of the multiple-track vertex using the following function:
\begin{equation}
\Ri^{\rm multiple}(\delta \zi) = G(\delta \zi; (\si^0+\si^1\xi)\sigma^\zi) \, \ (q={\rm ful,\ asc}),
\end{equation}
where 
$G$ is the Gaussian function,
\begin{equation}
G(x;\sigma) \equiv \frac{1}{\sqrt{2\pi}\sigma}\exp\left( -\frac{x^2}{2\sigma^2} \right).
\end{equation}
The scale factors $\si^0$ and $\si^1$ are treated as free parameters 
and determined from the  lifetime fit  to the data.
%   \begin{figure}
%     \makebox[0.45\textwidth][l]{(a)} \hspace*{0.05\textwidth}
%     \makebox[0.45\textwidth][l]{(b)}
%     \resizebox{\textwidth}{!}{\includegraphics{fig/vertex-residual.eps}}
%     \vspace*{\baselineskip}\\
%   \begin{center}
%     \makebox[0.45\textwidth][l]{(c)}\\
%     \resizebox{0.5\textwidth}{!}{\includegraphics{fig/convol-residual.eps}}
%   \end{center}
%     \caption{\label{fig:rdet_fit} Distributions of (a) $\delta \zrec$ and (b) $\delta \zasc$
%       for multiple-track vertices,  with $\Rrec^{\rm multiple}(\delta\zrec)$ and
%       $\Rasc^{\rm multiple}(\delta\zasc)$, respectively.
%       Figure (c) is the  $\delta(\delz)$ distribution,
%       and the convolution of
%       $\Rrec^{\rm multiple}(\delta\zrec)$ and
%       $\Rasc^{\rm multiple}(\delta\zasc)$.}
%   \end{figure}
\begin{figure}
  \resizebox{\textwidth}{!}{\includegraphics{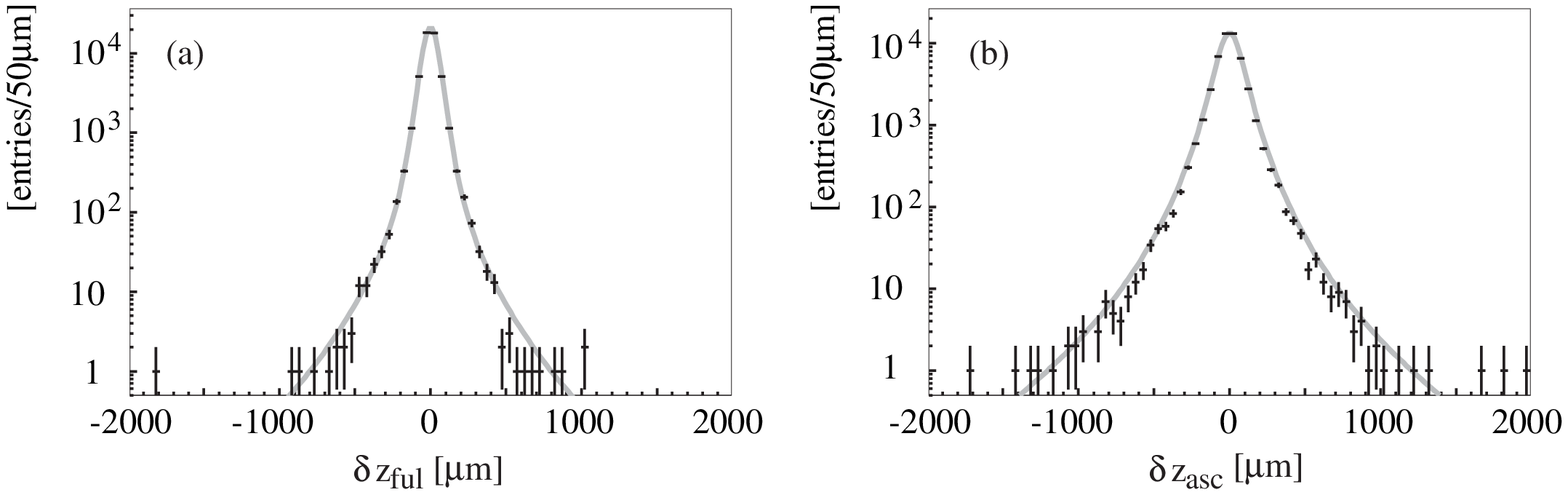}}
\begin{center}
  \resizebox{0.5\textwidth}{!}{\includegraphics{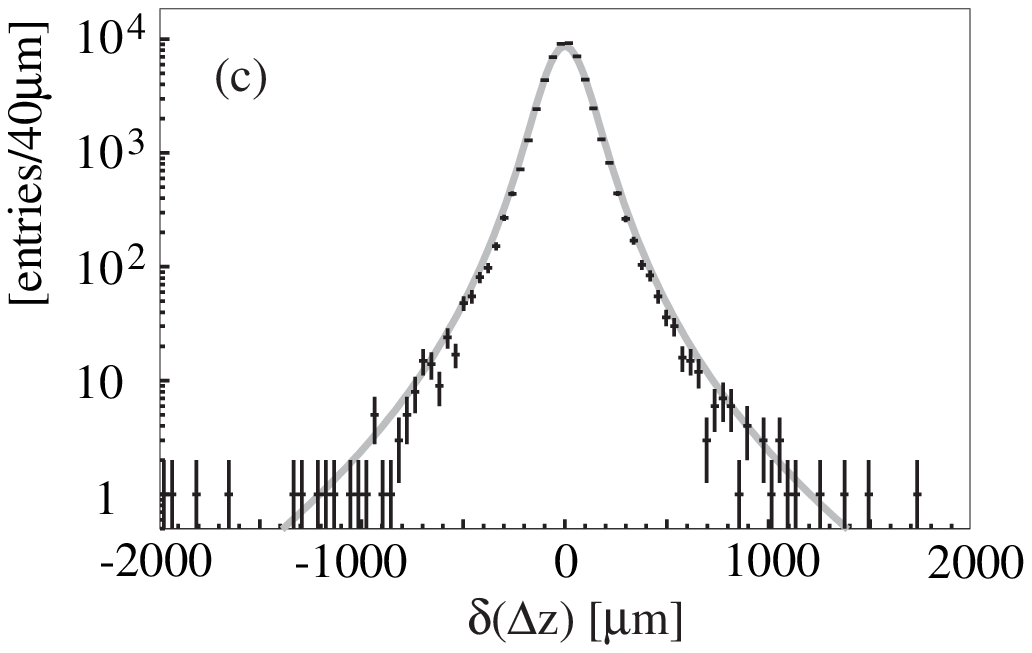}}
\end{center}
  \caption{\label{fig:rdet_fit} Distributions of (a) $\delta \zrec$ and (b) $\delta \zasc$
    for multiple-track vertices,  with $\Rrec^{\rm multiple}(\delta\zrec)$ and
    $\Rasc^{\rm multiple}(\delta\zasc)$, respectively.
    Figure (c) is the  $\delta(\delz)$ distribution,
    and the convolution of
    $\Rrec^{\rm multiple}(\delta\zrec)$ and
    $\Rasc^{\rm multiple}(\delta\zasc)$.}
\end{figure}
Figures~\ref{fig:rdet_fit} (a) and (b) show the $\delta z_{\rm ful}$  and  $\delta z_{\rm asc}$ distributions,
respectively.
%Superimposed are the results of a fit to $\Ri^{\rm multiple}(\delta \zi)$, which well
%reproduce the $\delta \zi$ distributions.
Superimposed are the curves obtained by summing vertex-by-vertex $\Ri^{\rm multiple}(\delta \zi)$
functions.
The curves well reproduce the $\delta \zi$ distributions.
This demonstrates that $\Ri^{\rm multiple}$ represents the detector resolution
better than the sum of two Gaussians in Fig.~\ref{fig:poor}.  
Figure \ref{fig:rdet_fit} (c) shows  the distribution of the residual of $\delz$, 
$\delta(\delz)\equiv \delz^{\rm rec} -\delz^{\rm gen}$ together with the convolution of $\Rrec^{\rm multiple}(\delta\zrec)$ and
$\Rasc^{\rm multiple}(\delta\zasc)$.

\subsubsection{Single-track vertex}
\begin{figure}
  \resizebox{\textwidth}{!}{\includegraphics{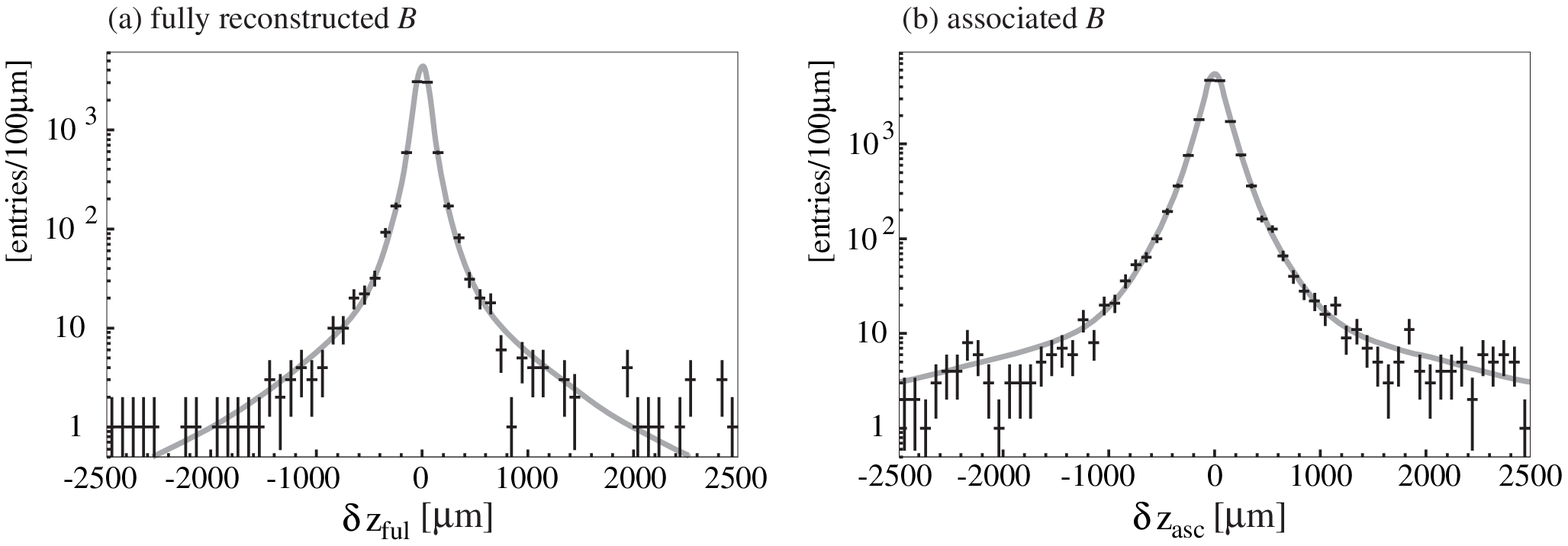}}
  \caption{\label{fig:single} The  (a) $\delta\zrec$
    and (b) $\delta\zasc$ distributions for single-track vertices with
     $\Rrec^{\rm single}$ and $\Rasc^{\rm single}.$}
\end{figure}

For the single-track vertices, $\xi$ is not available.
The resolution function of the single-track vertices,
$\Ri^{\rm single}(\delta\zi)\  (q={\rm ful,\ asc})$, is expressed as 
a sum of two Gaussians, 
one for the main part of the detector resolution 
and the other for the tail part, which is due to poorly reconstructed tracks:
\begin{equation}
  \Ri^{\rm single}(\delta\zi) = (1-\ftail) G(\delta\zi; \smain\Si^z)
  + \ftail G(\delta\zi; \stail\Si^z),
\end{equation}
where $\smain$ and $\stail$ are global scale factors which are common to all
single-track vertices.
Figure \ref{fig:single} shows the residual distributions of the single-track $\zrec$ and
$\zasc$ vertices, together with a fit to $\Ri^{\rm single}(\delta\zi)$. 
%   \begin{figure}
%     \makebox[0.45\textwidth][l]{(a)} \hspace*{0.05\textwidth}
%     \makebox[0.45\textwidth][l]{(b)}
%     \resizebox{\textwidth}{!}{\includegraphics{fig/vertex-residual-single.eps}}
%     \caption{\label{fig:single} The  (a) $\delta\zrec$
%       and (b) $\delta\zasc$ distributions for single-track vertices with
%        $\Rrec^{\rm single}$ and $\Rasc^{\rm single}.$}
%   \end{figure}

\subsection{Smearing due to non-primary tracks}
\begin{figure}
\begin{center}
{\resizebox{0.5\textwidth}{!}{\includegraphics{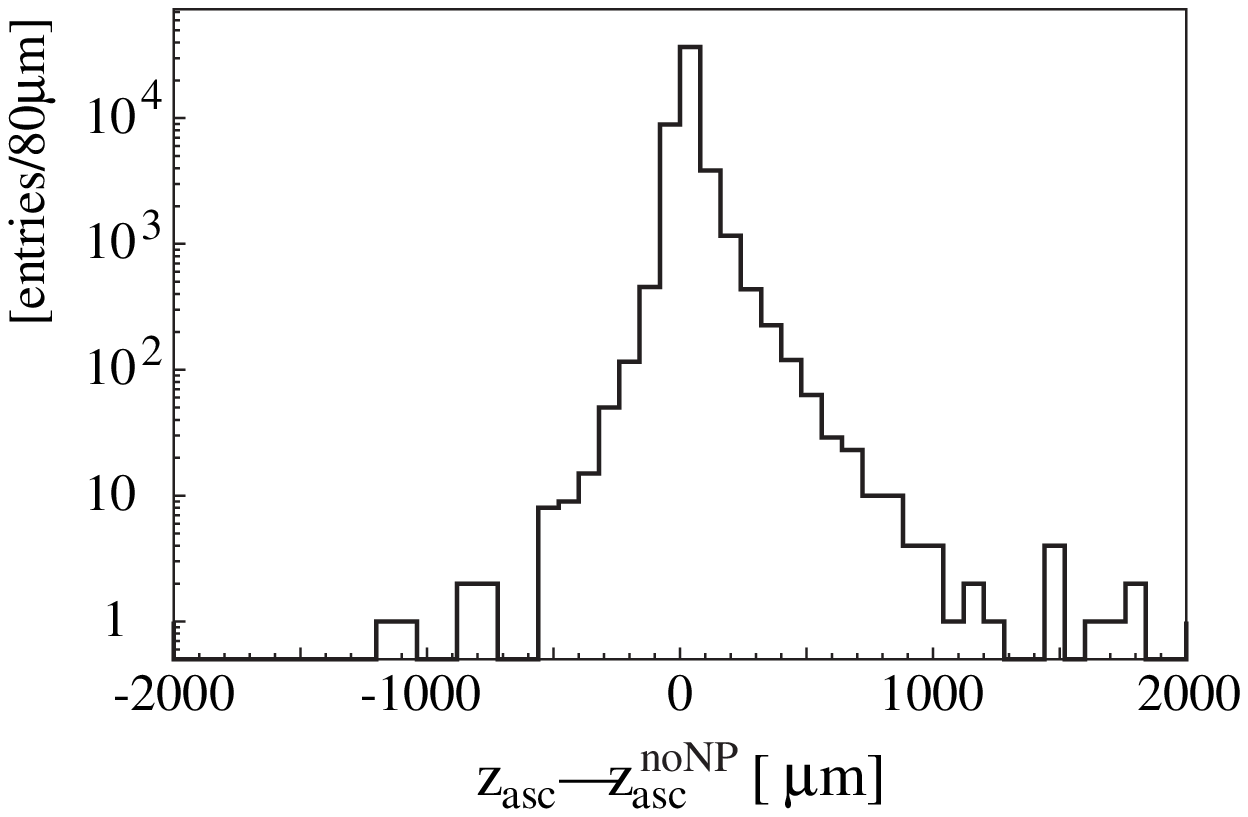}}}
  \caption{  \label{fig:charm}Distribution of
    $\znorm - \znonp$ for multi-track vertices, where $\znonp$ is $\zasc$ obtained from a MC
sample in which secondary decays are turned off. 
In making this plot the events in which $\znonp = \znorm$ are removed.
	The histogram is obtained from $\bzbpsiks$ MC 
	whose associated $B$ vertex is reconstructed with multiple tracks.
	}
\end{center}
\end{figure}

\begin{figure}
  % \makebox[0.45\textwidth][l]{(a)} \hspace*{0.05\textwidth}
  % \makebox[0.45\textwidth][l]{(b)}
  \resizebox{\textwidth}{!}{\includegraphics{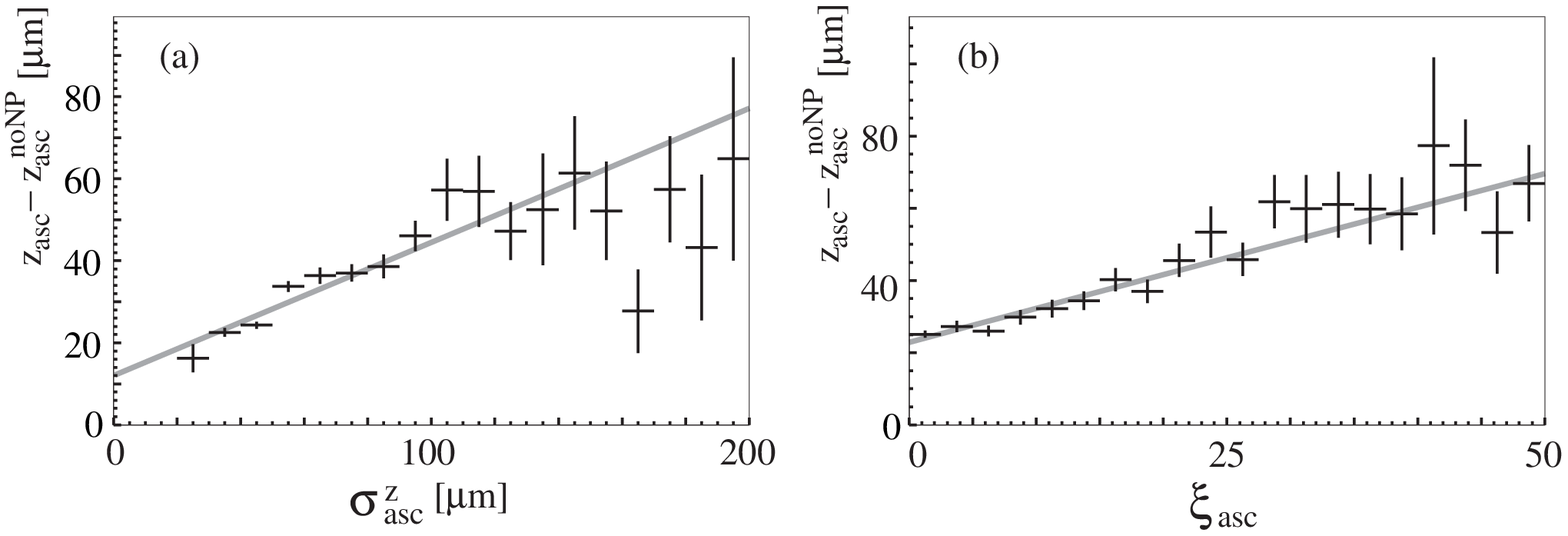}}
  \caption{ \label{fig:chmz_vs}  Average shift of vertex position  ($\znorm-\znonp$) vs
    (a) $\Sasc$ and (b) $\xi_{\rm asc}$. $\znonp$ is obtained from a MC
sample in which secondary decays are turned off.
    The events with $\znorm < \znonp$ are excluded.}
\end{figure}

We introduce another resolution function,
$\Rnp$, to represent the smearing of $\zasc$ due to tracks that  do not
originate from the associated $B$ vertex 
consists of  a prompt component, expressed by Dirac's $\delta$-function
$\delta^{\rm  Dirac}(\delta\zasc)$, and components that  account for smearing due to
$\ks$ and charm decays.
The functional form of the non-prompt components
is determined from the difference between  
$\zasc$ obtained for  the nominal MC sample and that for the special MC sample 
in which all short-lived secondary particles are forced to decay with zero lifetime
at the $B$ decay points, shown in Fig.~\ref{fig:charm}.
It can be expressed  by a function defined as
$\fp \Ep(\delta\zasc; \taunpp) +
    (1-\fp) \En(\delta\zasc; \taunpn),
$
where $\fp$ is a fraction of the $\delta\zasc>0$ component and 
$\Ep$ and $\En$ are:
\begin{eqnarray}
\label{eq:exponential}
  \Ep(x;\tau) &\equiv& \frac{1}{\tau} \exp\left(-\frac{x}{\tau}\right)
  {\rm  for}\  x > 0,  {\rm otherwise}\  0 , \\
  \En(x;\tau) &\equiv& \frac{1}{\tau} \exp\left(+\frac{x}{\tau}\right)
  {\rm  for}\  x\le 0,\ {\rm otherwise}\ 0.
\end{eqnarray}
Thus, $\Rnp$ is given by 
\begin{equation}
  \label{eq:rcharm}
  \Rnp(\delta\zasc) \equiv f_\delta \delta^{\rm Dirac}(\delta\zasc) +
  (1-f_\delta) \left[
    \fp \Ep(\delta\zasc; \taunpp) +
    (1-\fp) \En(\delta\zasc; \taunpn) \right],
\end{equation}
where $f_\delta$ is the prompt-component fraction.
We find  that the vertex position shift 
has linear dependence on both
$\sigma^z_{\rm asc}$ and $\xi^z_{\rm asc}$ as shown in
Fig.~\ref{fig:chmz_vs}  for multi-track vertices.
Consequently, we assume  $\taunpp$ and $\taunpn$ 
are bilinear with $\sigma_{\rm asc}$ and $\xi_{\rm asc}$ as
\begin{eqnarray}
  \taunpp &=& \taup^0 + \taup^1
  (\sasc^0 + \sasc^1 \xi^z_{\rm asc}) \Sasc/c(\beta\gamma)_\Upsilon ,\ \ {\rm and} \\
  \taunpn &=& \taun^0 + \taun^1
  (\sasc^0 + \sasc^1 \xi_{\rm asc}) \Sasc/c(\beta\gamma)_\Upsilon .
\end{eqnarray}

\begin{figure}
  % \makebox[0.45\textwidth][l]{(a)} \hspace*{0.05\textwidth}
  % \makebox[0.45\textwidth][l]{(b)}
  \resizebox{\textwidth}{!}{\includegraphics{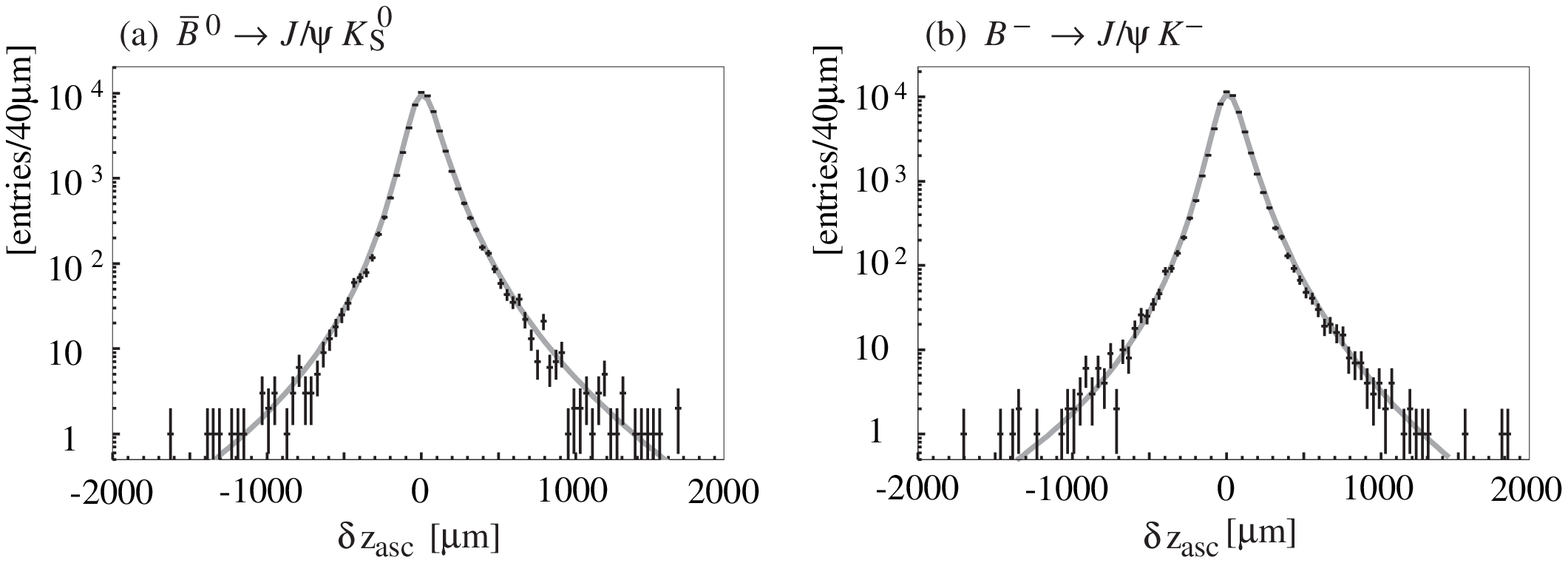}}
  \caption{\label{fig:rcharm_mul} 
The   $\delta\zasc$ distributions of  multiple-track vertices for
  (a) $\bzbpsiks$ and (b) $\bmpsikm$ decays.}
\end{figure}

We determine six parameters in $\Rnp$,  $f_\delta$, $\fp$, $\taup^0$, $\taup^1$, $\taun^0$,
and  $\taun^1$ by fitting the convolution of  $\Rasc^{\rm multiple}$ and $\Rnp$ to  the $\delta\zasc$ distributions
for $\overline{B}{}^0$ and $B^-$ separately,
as shown in Fig.~\ref{fig:rcharm_mul}. 
In this fit, the scale parameters, $s^0_{\rm asc}$ and $s^1_{\rm asc}$,  for $\Rasc^{\rm multiple}$ are fixed to 
the values (common to $\overline{B}{}^0$ and $B^-$) obtained by fitting $\Rasc^{\rm multiple}$ to 
the special MC sample (in which all short-lived secondary particles  are forced to decay promptly
at the $B$ meson decay points).
Results, shown superimposed, well represent the distributions.

\begin{figure}
\begin{center}
  \resizebox{0.5\textwidth}{!}{\includegraphics{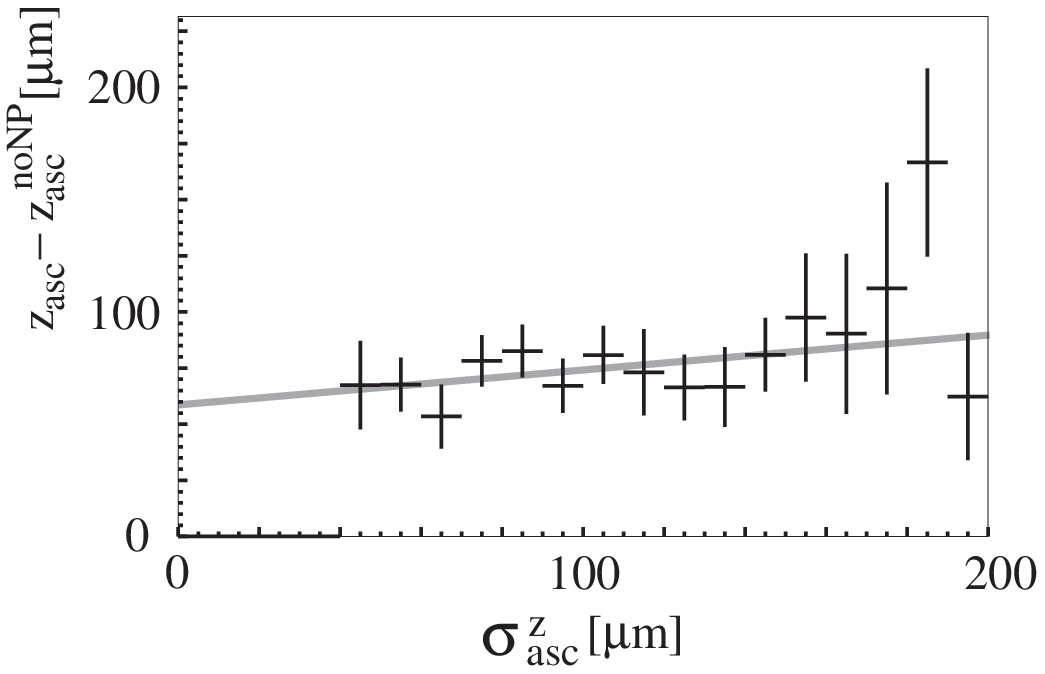}}
\end{center}
  \caption{\label{fig:chmz_vs_single} Average shift of vertex positions  ($\znorm-\znonp$) versus
  $\Sasc$ for single-track vertices.
    The events where $\znorm < \znonp$ are excluded.}
\end{figure}

For single-track vertices we  consider the correlation between the vertex
position shift and $\Sasc$.  
Figure \ref{fig:chmz_vs_single} shows the vertex position shift versus $\Sasc$ for the single-track vertices. 

\begin{figure}
  % \makebox[0.45\textwidth][l]{(a)} \hspace*{0.05\textwidth}
  % \makebox[0.45\textwidth][l]{(b)}
  \resizebox{\textwidth}{!}{\includegraphics{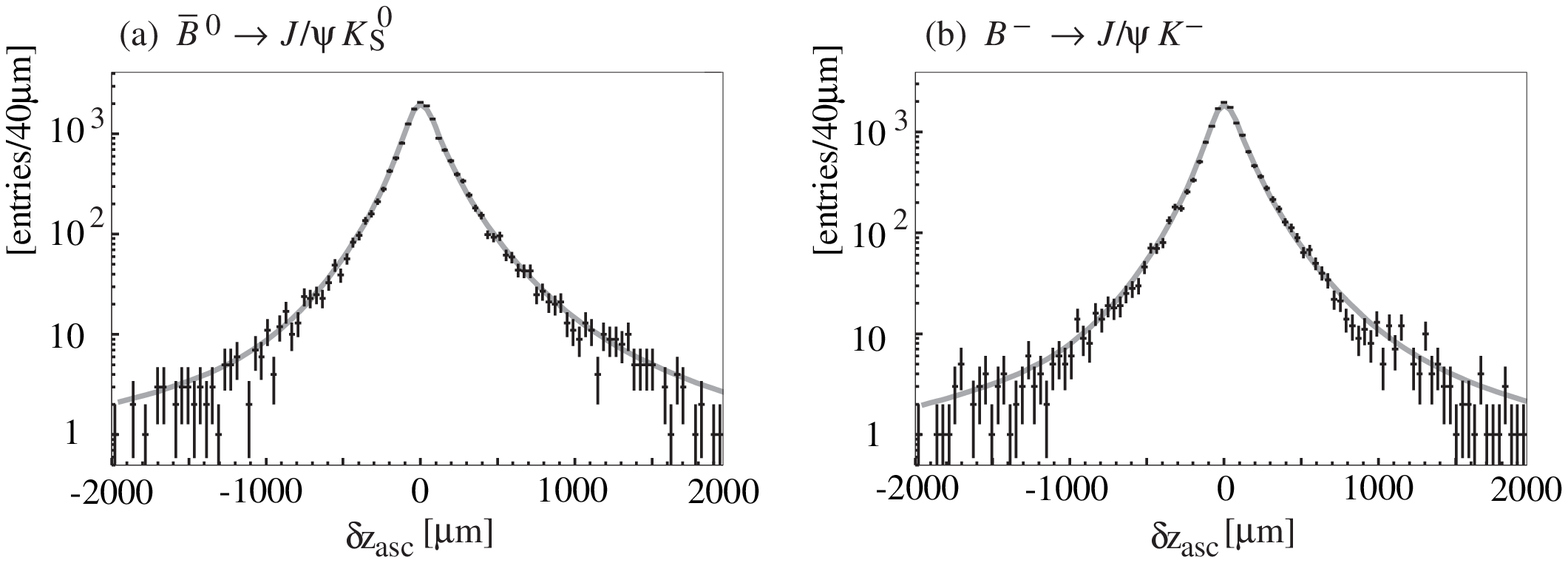}}
  \caption{\label{fig:rcharm_single} The $\delta \zasc$ distributions 
    of single-track vertices 
    for (a) $\overline{B}{}^0$ and (b) $B^-$ mesons.}
\end{figure}

Since $\Rasc$ for the single-track vertices is defined as a sum of main and tail Gaussians,
we also introduce $\Rnp^{\rm main}$ and $\Rnp^{\rm tail}$ for main and tail parts,
respectively. 
Each of $\Rnp^{\rm main}$ and $\Rnp^{\rm tail}$ is expressed by the function of
Eq.~(\ref{eq:rcharm}) with parameters defined as:
\begin{eqnarray}
    (\taunpp)_{\rm main}& =& \taup^0 + \taup^1 \smain \Sasc/c(\beta\gamma)_\Upsilon \nonumber\\
    (\taunpn)_{\rm main}& = &\taun^0 + \taun^1 \smain \Sasc/c(\beta\gamma)_\Upsilon, \\
    (\taunpp)_{\rm tail\ \ } &=& \taup^0 + \taup^1 \stail \Sasc/c(\beta\gamma)_\Upsilon, \ \ {\rm and}\nonumber \\
    (\taunpn)_{\rm tail\ \ } &=& \taun^0 + \taun^1 \stail \Sasc/c(\beta\gamma)_\Upsilon .
\end{eqnarray}
The convolution of $\Rasc$ and $\Rnp$ for single-track vertices is, thus, defined as:
\begin{eqnarray}
\label{eq:rcharm_single}
  \Rasc^{\rm single} \otimes \Rnp^{\rm single} (\delta\zasc) &=& 
  \int_{-\infty}^{+\infty} \mathrm{d}(\delta\zasc^\prime) 
  \left[ (1-\ftail) G(\delta\zasc-\delta\zasc^\prime; \smain\Si) 
 \Rnp^{\rm main}(\delta\zasc^\prime) \right] 
  \nonumber \\
  &+& \int_{-\infty}^{+\infty} \mathrm{d}(\delta\zasc^\prime)
  \left[ \ftail G(\delta\zasc-\delta\zasc^\prime; \stail\Si) \Rnp^{\rm tail}(\delta \zasc^\prime) \right] .
\end{eqnarray}
Figure~\ref{fig:rcharm_single} shows the $\delta \zasc$ distributions for the single track vertices.
The superimposed curves are the results of a fit to the function given by
Eq.~(\ref{eq:rcharm_single}).

Table~\ref{tab:Rnp} lists the shape parameters of $\Rnp$ determined 
by fitting $\Rasc\otimes \Rnp(\delta\zasc)$ to MC $\delta\zasc$ distributions.
These parameter values are held  fixed when the lifetime fit to the data is performed.

%LP01
\begin{center}
\begin{table}
  \caption{\label{tab:Rnp} Shape  parameters of $\Rnp$ used for the lifetime fit.
The values are  determined for multiple- and single-track
vertices, separately,  using  Monte Carlo $\delta\zasc$ distributions.}
    \begin{tabular}{lccccc}
\hline
\hline
      & \multicolumn{2}{c}{$\bzb$} & & \multicolumn{2}{c}{$\bm$} \\
      & multiple & single & & multiple & single \\
      \hline
      $f_\delta$      & $0.676\pm0.007$            & $0.787\pm 0.011$  & & $0.650\pm0.010$   & $0.763\pm 0.018$  \\
      $\fp$           & $0.955\pm0.004$            & $0.790\pm 0.021$  & & $0.963\pm0.004$       & $0.757\pm 0.026$  \\
      $\taup^0$ (ps ) & $-0.010\pm0.011$           & $0.108\pm0.068$  & & $0.037\pm0.012$  & $-0.019\pm0.066$ \\
      $\taup^1$       & $0.927\pm 0.025$  & $1.321\pm  0.097$  & & $0.674\pm0.025$           & $1.113\pm 0.096$  \\
      $\taun^0$ (ps)  & $-0.194\pm 0.078$ & $-0.281_{-0.147}^{+0.130}$ & & $-0.269\pm0.099$ & $-0.375_{-0.122}^{+0.111}$ \\
      $\taun^1$       & $1.990_{-0.169}^{+0.182}$  & $1.583_{-0.184}^{+0.213}$  & & $2.070_{-0.213}^{+0.235}$ & $1.548_{-0.182}^{+0.207}$  \\
\hline
\hline
    \end{tabular}
\end{table}
\end{center}

\subsection{Kinematic approximation}

\begin{figure}
\begin{center}
  \resizebox{0.5\textwidth}{!}{\includegraphics{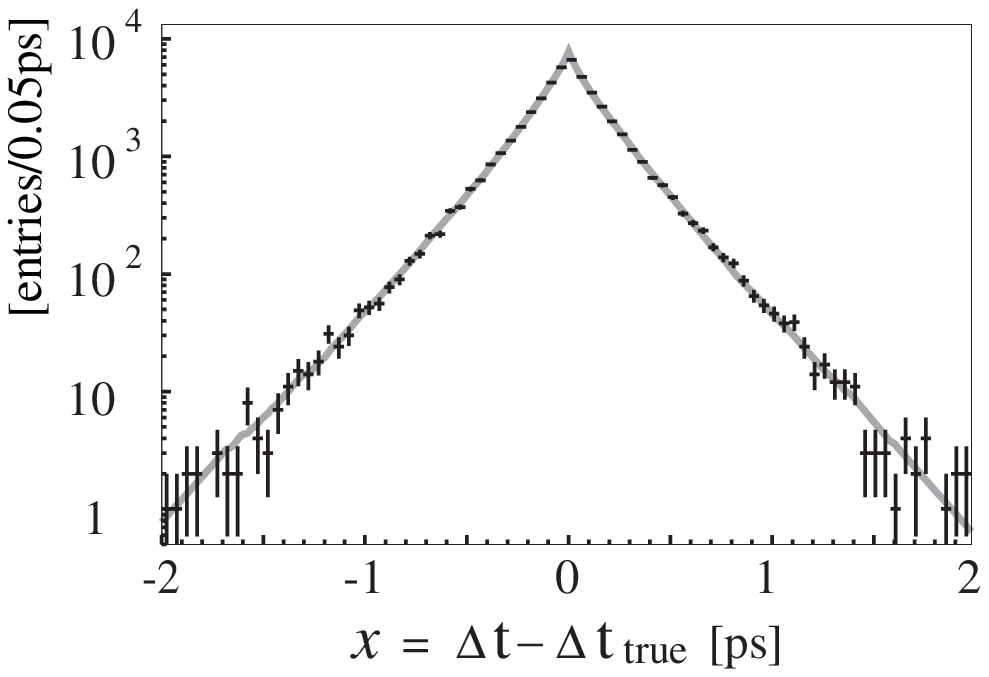}}
\end{center}
  \caption{\label{fig:rkfig} 
  The $x=\Delta t - \Delta t_{\rm true}$ distribution for $\overline{B}{}^0\to J/\psi \ks$ events
together with the function $\Rk(x)$.}
\end{figure}

The proper time interval calculated as Eq.~(\ref{eq;deltat}) , $\Delta t=(\zrec-\zasc)/[c(\beta\gamma)_\Upsilon]$, is equal to the {\it true} proper-time interval when the cms motion 
of the $B$ mesons is neglected.
The difference between $\Delta t$ and the true proper-time interval $\Delta t_{\rm true}=\trec-\tasc$ is calculated from the kinematics of the $\Upsilon(4S)$ two-body decay:
\begin{eqnarray}
x\equiv  \Delta t - \Delta t_{\rm true} &= &(\zrec - \zasc)/[c\bgu]-(\trec-\tasc)\nonumber \\
 &=& [\trec c\bgrec - \tasc c\bgasc]/[c\bgu] - (\trec-\tasc)\nonumber \\
 &=& [\bgrec/\bgu -1]\trec - [\bgasc/\bgu - 1]\tasc,
\end{eqnarray}
where $\bgrec$ and $\bgasc$ are Lorentz boost factors of the fully reconstructed and
associated $B$ mesons, respectively, and their ratios to $\bgu$ are given as:
\begin{eqnarray}
  (\beta\gamma)_{\rm ful}/\bgu &=&
 \frac{\Eb}{m_B} + \frac{\pb\cosb}{\beta_\Upsilon m_B}
 \sim  1+0.165\cosb, \ \ {\rm and}\\
   (\beta\gamma)_{\rm asc}/\bgu &=&  \frac{\Eb}{m_B}
    - \frac{\pb\cosb}{\beta_\Upsilon m_B}\sim 1-0.165\cosb,
\end{eqnarray}
where $\beta_\Upsilon=0.391$ is the velocity of the $\Upsilon(4S)$ in units of $c$,
$E_B^{\rm cms}\sim 5.292~{\rm GeV}$, $p_B^{\rm cms}\sim 0.340~{\rm GeV}/c$ 
and $\theta_B^{\rm cms}$ are the energy, momentum and  
polar angle of the fully reconstructed $B$ in the cms, and $m_B$ is either
the $\overline{B}{}^0$ or the $B^-$ mass.
The difference $x$ can, therefore, be approximated as
\begin{equation}
x\sim 0.165\cosb(\trec+\tasc).
\end{equation}
$\Rk$, which  accounts for $x$, can be  given as a function of $\cos\theta_B^{\rm cms}$.
Because $\trec$ and $\tasc$ distributions follow $\Ep(\trec;\tau_B)=\frac{1}{\tau_B}\exp(-\trec/\tau_B)$ and
$\Ep(\tasc;\tau_B)=\frac{1}{\tau_B}\exp(-\tasc/\tau_B)$, respectively, 
the probability density of obtaining $x$ and $\Delta t_{\rm true}$ simultaneously 
is given by:
\begin{eqnarray}
F(x,\Delta t_{\rm true})& = & \int_{0}^{\infty}\int_{0}^{\infty}\mathrm{d}\trec\mathrm{d}\tasc
\Ep(\trec ; \tau_B)\Ep(\tasc ; \tau_B)\delta^{\rm Dirac}(\Delta t_{\rm true}-(\trec-\tasc))\nonumber \\
&&\times\delta^{\rm Dirac}(x- \{[\bgrec/\bgu -1]\trec - [\bgasc/\bgu - 1]\tasc\}),\nonumber\\
\end{eqnarray}
and the probability density of obtaining $\Delta t_{\rm true}$ is given by:
\begin{equation}
F(\Delta t_{\rm true})= \int_{0}^{\infty}\int_{0}^{\infty}\mathrm{d}\trec\mathrm{d}\tasc
\Ep(\trec ;\tau_B)\Ep(\tasc; \tau_B)\delta^{\rm Dirac}(\Delta t_{\rm true}-(\trec-\tasc)).
\end{equation}
$\Rk(x)$ is, then,  defined as the conditional probability density of obtaining $x$ given $\Delta t_{\rm true}$.
It is expressed as $\Rk(x)=F(x,\Delta t_{\rm true})/F(\Delta t_{\rm true})$ which gives:
\begin{multline}
  \label{eq:rkend}
  \Rk(x) = 
  \begin{cases}
    \Ep\left(x -
      \Bigl[(\frac{\Eb}{m_B}-1)\dt_{\rm true} + \frac{\pb\cosb}{\beta_\Upsilon m_B}|\dt_{\rm true}|\Bigr];
      |\frac{\pb\cosb}{\beta_\Upsilon m_B}| \tau_B \right) & (\cos\theta_B^{\rm cms}> 0) \\
    \delta^{\rm Dirac}\left(x -
      (\frac{\Eb}{m_B}-1)\dt_{\rm true} \right) & (\cos\theta_B^{\rm cms} = 0) \\
    \En\left(x -
      \Bigl[(\frac{\Eb}{m_B}-1)\dt_{\rm true} + \frac{\pb\cosb}{\beta_\Upsilon m_B}|\dt_{\rm true}|\Bigr];
      | \frac{\pb\cosb}{\beta_\Upsilon m_B}|\tau_B\right) & ( \cos\theta_B^{\rm cms}< 0) 
  \end{cases}.
\end{multline}
Figure~\ref{fig:rkfig}  shows the $x$ distribution for $\overline{B}{}^0\to J/\psi \ks$ events
with the function $\Rk(x)$.
The expected  theoretical $\Delta t$ distribution $P(\Delta t)$ can be  expressed as
a convolution of the true PDF $\calPsig(\dt_{\rm true}; \tau_B)$  with  $\Rk(\Delta t - \Delta t_{\rm true})$:
\begin{equation}
  \label{eq:example_rk}
  P(\Delta t) = \frac{m_B}{2 \Eb \tau_B} \exp \left(
    -\frac{|\Delta t|}{ (\frac{\Eb}{m_B} \pm  \frac{\pb\cosb}{\beta_\Upsilon m_B }) \tau_B} \right)\qquad
\begin{cases}
    +\ {\rm for}\ \Delta t \ge 0\\
     -\ {\rm for}\ \Delta t < 0
  \end{cases}.
\end{equation}

\subsection{Background distribution}
The signal purity $f_{\rm sig}$ in Eq.~(\ref{eq:pdf}) is determined 
on an event-by-event basis as a function of two kinematic variables,
the energy difference $\Delta E=E_B^{\rm cms}-E_{\rm beam}^{\rm cms}$ and the beam-energy constrained
mass $M_{\rm bc}=\sqrt{(E_{\rm beam}^{\rm cms})^2-(p_B^{\rm cms})^2}$, where $E_{\rm beam}^{\rm cms}$
is the beam energy in the cms.
Typical $f_{\rm sig}$ values for the decay modes used for the measurement of $B$ meson lifetimes and 
$B^0\overline{B}{}^0$  oscillation frequency are listed in Table~\ref{tab:fsig}.
The values listed are obtained for the signal region defined as $|M_{\rm bc}-m_B|<3\sigma.$ 
The composition of the background is studied by a MC simulation and is listed in Table~\ref{tab:bkgcom}.
The largest contribution is from  $c\bar{c}$ events produced in the continuum,
 while other continuum events ($u,d,s$ events) and combinatorial backgrounds from $B\overline{B}$ events
also contribute.
\begin{table}
\caption{\label{tab:fsig}Typical values of the signal purity $f_{\rm sig}$ 
for the decay modes used for the measurement of $B$ meson lifetimes and 
$B^0\overline{B}{}^0$  oscillation frequency.}
\begin{center}
\begin{tabular}{l|c}
\hline
\hline
Decay mode $\Brec$ & Signal purity $f_{\rm sig}$ \\
\hline
$\overline{B}{}^0\rightarrow D^+\pi^-$   & 0.861\\
$\overline{B}{}^0\rightarrow D^{*+}\pi^-$  & 0.812\\
$\overline{B}{}^0\rightarrow D^{*+}\rho^-$ & 0.699\\
$\overline{B}{}^0\rightarrow J/\psi \ks$  &0.940\\
$\overline{B}{}^0\rightarrow J/\psi K^{*0}$ &0.948\\
$B^-\rightarrow D^0\pi^-$    &0.699\\
$B^-\rightarrow J/\psi K^-$  &0.955\\
\hline
\hline
\end{tabular}
\end{center}
\end{table}
\begin{table}
\caption{\label{tab:bkgcom}The composition of the background in 
the decay modes used for the measurement of $B$ meson lifetimes and 
$B^0\overline{B}{}^0$  oscillation frequency. The sources of the backgrounds are  the continuum
production of $u\bar{u},d\bar{d},s\bar{s}$ pairs ($q\bar{q}(u,,d,s)$) and $c\bar{c}$ pairs ($q\bar{q}(c))$, and
 combinatorial backgrounds from $B^+B^-$ and $B^0\overline{B}{}^0$ events.
$\overline{B}\rightarrow J/\psi X$ modes, in which the background fraction is very small, are not listed.}
\begin{center}
\begin{tabular}{l|cccc}
\hline
\hline
Decay mode $\Brec$&  $q\bar{q}(u,d,s)$  &  $q\bar{q}(c)$ &  $B^+B^-$  & $B^0\overline{B}{}^0$\\
\hline
$\overline{B}{}^0\rightarrow D^+\pi^-$ &  0.25 & 0.40 & 0.21& 0.14\\
$\overline{B}{}^0\rightarrow D^{*+}\pi^-$&  0.10 &  0.48&   0.24&  0.18\\
$\overline{B}{}^0\rightarrow D^{*+}\rho^-$&0.10&  0.43&  0.26&  0.21\\
$B^-\rightarrow D^0\pi^-$  &0.29 &  0.45&  0.17&   0.09\\
\hline
\hline
\end{tabular}
\end{center}
\end{table}

The background PDF, $\Pbg(\Delta t)$, is modeled as a sum of exponential and 
prompt components ($\mathcal{P}_{\rm bkg}(\Delta t)$) convolved with $\Rbg(\Delta t)$:
\begin{equation}
\Pbg(\Delta t)=\int^{+\infty}_{-\infty} {\rm d}(\Delta t^\prime)\mathcal{P}_{\rm bkg}(\Delta t^\prime)
\Rbg(\Delta t - \Delta t^\prime),
\end{equation}
where
\begin{equation}
\mathcal{P}_{\rm bkg}(\Delta t)= f_\delta^{\rm bkg}\delta^{\rm Dirac}(\Delta t -\mu_\delta
^{\rm bkg})+(1-f_\delta^{\rm bkg})\frac{1}{2\tau_{\rm bkg}}\exp\left(-\frac{|\Delta t-\mu
_\tau^{\rm bkg}|}{\tau_{\rm bkg}}\right)
\end{equation}
with $\mu_\delta^{\rm bkg}$ and $\mu_\tau^{\rm bkg}$ being offsets of the distribution, and
\begin{equation}
\Rbg(\Delta t)=(1-f_{\rm tail}^{\rm bkg})G(\Delta t;s_{\rm main}^{\rm bkg}\sqrt{\sigma_{\rm ful}^2
+\sigma_{\rm asc}^2})+f_{\rm tail}^{\rm bkg}G(\Delta t; s_{\rm tail}^{\rm bkg}\sqrt{\sigma_{\rm ful}^2
+\sigma_{\rm asc}^2}).
\end{equation}
Different  values are used for $s_{\rm main}^{\rm bkg}$, $s_{\rm tail}^{\rm bkg}$, and $f_{\rm tail}^{\rm bkg}$
depending on whether both vertices are reconstructed with multiple tracks or not.
All parameters  in $\Pbg(\Delta t)$ are determined by the fit to the $\Delta t$ distribution of
the background-enhanced control sample ({\it i.e.}~events in the sideband region of the $\Delta E$ or $M_{\rm bc}$
distribution).
Table~\ref{tab:bkgshape} lists the values obtained separately for each  decay mode 
used in the lifetime fit.

\begin{center}
\begin{table}
  \caption{\label{tab:bkgshape} The background shape parameters obtained by fitting $P_{\rm bkg}(\Delta t)$
to the background-enhanced control sample.}
 \begin{tabular}{lccc}
(a) $\overline{B}\to J/\psi \overline{K}$ modes &&&\\
\hline
\hline
      Parameter                    & $\bzbpsiks$             & $\bzbpsikst$            & $\bmpsikm$              \\
\hline
      $(\smainbg)_\text{multiple}$ & $ 0.40\pm 0.12$ & $ 1.09\pm 0.25$ & $ 0.79\pm 0.23$ \\
      $(\stailbg)_\text{multiple}$ & $ 9.46^{+4.26}_{-2.57}$ & $ 6.97^{+6.39}_{-2.30}$ & $ 1.90^{+0.64}_{-0.37}$ \\
      $(\ftailbg)_\text{multiple}$ & $ 0.29\pm 0.12$ & $ 0.03\pm 0.04$ & $ 0.66^{+0.21}_{-0.28}$ \\
      $(\fdelbg)_\text{multiple}$  & $ 0.39\pm 0.18$ & $ 0.08 \pm 0.08$        & $ 0.85\pm 0.05$ \\
      $(\smainbg)_\text{single}$   & $ 0.96^{+0.19}_{-0.23}$ & $ 0.82\pm0.15$ & $ 1.03\pm 0.10$ \\
      $(\stailbg)_\text{single}$   & $ 4.90^{+2.15}_{-1.29}$ & $ 8.33^{+2.96}_{-2.13}$ & $ 11.2^{+4.7}_{-3.0}$   \\
      $(\ftailbg)_\text{single}$   & $ 0.16^{+0.16}_{-0.07}$ & $ 0.09\pm 0.04$ & $ 0.05\pm 0.03$ \\
      $(\fdelbg)_\text{single}$    & $ 0.38^{+0.28}_{-0.34}$ & $ 0.18\pm 0.13$ & $ 0.65\pm 0.10$ \\
      $\tbg$ (ps)                  & $ 0.39\pm0.25$ & $ 1.43 \pm 0.16$        & $ 2.14^{+0.41}_{-0.33}$ \\
      $\mudelbg$ (ps)              & $-0.56 \pm 0.09$        & $-0.70\pm 0.28$ & $-0.00 \pm 0.05$        \\
      $\mutaubg$ (ps)              & $ 0.23\pm 0.18$ & $-0.00\pm 0.14$ & $-0.21^{+0.33}_{-0.37}$ \\
\hline
\hline
    \end{tabular}
  \vspace*{1cm}
    \begin{tabular}{lcccc}
  (b) $\bdpi$ modes&&&\\
\hline
\hline
      Parameter                    & $\bzdppm$               & $\bzdstpm$              & $\bzdstrhom$            & $\bmdzpm$               \\
      \hline
      $(\smainbg)_\text{multiple}$ & $ 1.03 \pm 0.04$        & $ 0.69^{+0.18}_{-0.14}$ & $ 0.90 \pm 0.07$        & $ 1.02 \pm 0.02$        \\
      $(\stailbg)_\text{multiple}$ & $ 3.03^{+0.68}_{-0.37}$ & $ 2.33^{+0.39}_{-0.32}$ & $ 5.19^{+0.74}_{-0.58}$ & $ 6.27\pm 0.36$ \\
      $(\ftailbg)_\text{multiple}$ & $ 0.14\pm 0.06$ & $ 0.67^{+0.12}_{-0.17}$ & $ 0.13\pm 0.04$ & $ 0.060 \pm 0.008$      \\
      $(\fdelbg)_\text{multiple}$  & $ 0.39\pm 0.09$ & $ 0.38 \pm 0.08$ & $ 0.22 \pm 0.07$ & $0.43\pm 0.03$         \\
      $(\smainbg)_\text{single}$   & $ 0.73 \pm 0.07$        & $ 0.87\pm 0.10$ & $ 0.93 \pm 0.08$        & $ 0.79 \pm 0.03$        \\
      $(\stailbg)_\text{single}$   & $ 4.69^{+0.90}_{-0.76}$ & $ 4.54^{+4.05}_{-1.34}$ & $ 3.52^{+0.47}_{-0.39}$ & $ 5.99\pm 0.54$ \\
      $(\ftailbg)_\text{single}$   & $ 0.12 \pm 0.04$        & $ 0.08\pm 0.05$ & $ 0.17 \pm 0.05$        & $ 0.09 \pm 0.01$        \\
      $(\fdelbg)_\text{single}$    & $ 0.29\pm 0.14$ & $ 0.32 \pm 0.18$        & $ 0.19 \pm 0.10$        & $0.30 \pm 0.05$         \\
      $\tbg$ (ps)                  & $ 1.10\pm 0.14$ & $ 1.68^{+0.26}_{-0.20}$ & $ 0.87\pm 0.11$ & $ 0.98 \pm 0.05$        \\
      $\mudelbg$ (ps)              & $-0.03 \pm 0.03$        & $ 0.00 \pm 0.03$        & $ 0.11\pm 0.07$ & $-0.02 \pm 0.01$        \\
      $\mutaubg$ (ps)              & $ 0.00\pm 0.08$  & $-0.11\pm 0.14$ & $-0.13\pm 0.07$ & $-0.11 \pm 0.02$        \\
\hline
\hline
    \end{tabular}
 \end{table}
\end{center}

\subsection{Outliers}
We find that there still exists a very long tail that cannot be described
by the resolution functions discussed above.  The outlier term is
introduced to describe this long tail and is represented by
a single Gaussian with zero mean and event-independent width:
\begin{equation}
  \label{eq:outlier}
  \Pol(\dt) = G(\dt, \sigol).
\end{equation}
The global fraction of outliers $\fol$ (in  Eq.~(\ref{eq:pdf})) 
and the $\sigol$ are left as free parameters in the lifetime fit. 
Different values are obtained for $\fol$ 
depending on whether both vertices are reconstructed with multiple tracks or not
($\fol^{\rm multiple}, \fol^{\rm single}$).
We find that $\fol^{\rm multiple}$ is less than $10^{-3}$ and $\fol^{\rm single}\sim 10^{-2}$.

%% file: application.tex
\section{Applications}
\label{sec:application}
\subsection{$B$ meson lifetimes and $B^0\overline{B}{}^0$ oscillation}
\begin{figure}
\begin{center}
  \resizebox{0.6\textwidth}{!}{\includegraphics{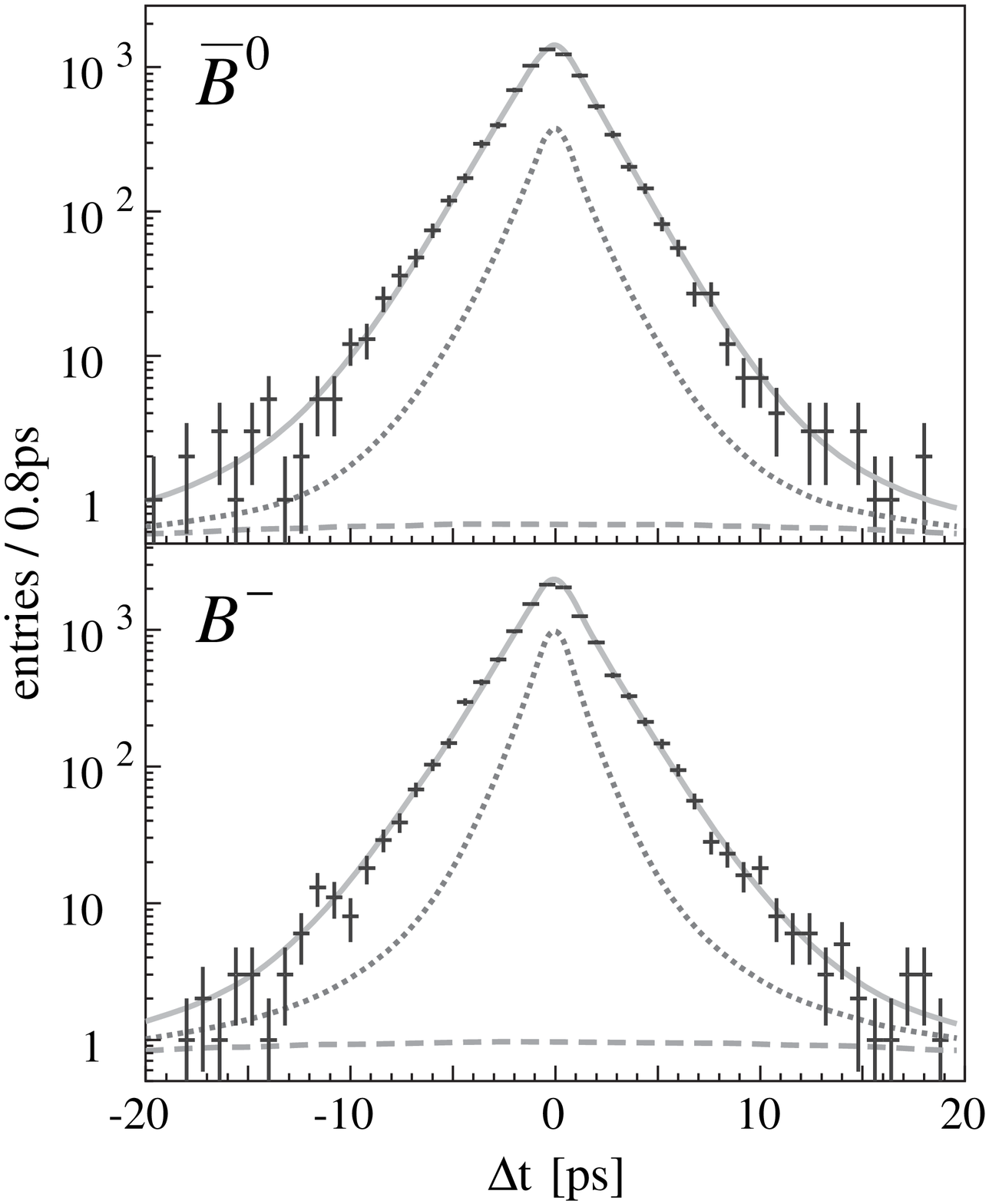}}%
\end{center}
  \caption{\label{fig:lifefit}
    The $\dt$ distributions of (a) neutral and (b) charged
    $B$ meson pairs, with fitted curves.  The dotted lines
    represent the sum of the background and outlier component,
    and the dashed lines represent the outlier component.}
\end{figure}

The lifetimes of the $\overline{B}{}^0$ and $B^-$ mesons are extracted from a 29~${\rm fb}^{-1}$ data sample,
which contains 31.3 million $B\overline{B}$ pairs~\cite{blife}.
In  the final fit to the events in the signal region, we determine simultaneously twelve parameters:
the $\overline{B}{}^0$ and $B^-$ lifetimes ($\taubz$, $\taubm$), four scale factors ($\srec^0, \srec^1, \sasc^0,
\sasc^1$) for $\Rrec^{\rm multiple}$ and $\Rasc^{\rm multiple}$, 
three parameters ($s_{\rm main}$, $s_{\rm tail}$ and $f_{\rm tail}$) for  
$\Rrec^{\rm single}$ and $\Rasc^{\rm single}$,  and three parameters ($\sigol$, $\fol^{\rm multiple}$,
$\fol^{\rm single}$) for outlier component.
Table~\ref{tab:lifefit} lists the result of the fit.
The fit yields:
\begin{eqnarray*}
  \taubz &=& \tbzresult~{\rm ps}, \\
  \taubm &=& \tbmresult~{\rm ps}, \ \ {\rm  and} \\
  \rbm &=& \rbmresult .
\end{eqnarray*}
The resulting
$\Delta t$ resolution for the signal is $\sim 1.56~{\rm ps}$ (rms) for this data sample.
Figure~\ref{fig:lifefit} shows the distributions of $\dt$ 
for $\bzb$ and $\bm$ events in the signal region
with the fitted curves superimposed. 
The systematic errors arising from the resolution functions are estimated by comparing the results obtained 
using a different functional form and by varying the resolution parameters within their errors.
They amount to $0.010$~ps, $ 0.009$~ps and $\ 0.006$ for $\taubz$, $\taubm$ and $\rbm$, respectively.
MC simulation studies show no bias 
arising from the resolution function in the obtained results.
\begin{table}
\begin{center}
\caption{\label{tab:lifefit} The parameters determined by the lifetime fit.}
 \begin{tabular}{clcc}
\hline
\hline
      & Parameters             & Values                                  & \\
      \hline
      & $\taubz$ (ps)          & $1.554 \pm 0.030$                       & \\
      & $\taubm$ (ps)          & $1.695 \pm 0.026$                       & \\
      \hline
      & $\srec^0$              & $0.809 \pm 0.148$              & \\
      & $\srec^1$              & $0.154 \pm 0.013$                       & \\
      & $\sasc^0$              & $0.753 \pm 0.065$              & \\
      & $\sasc^1$              & $0.064 \pm 0.005$                       & \\
      & $\smain$               & $0.647 ^{+0.074}_{-0.083}$              & \\
      & $\stail$               & $3.00 ^{+2.23}_{-0.99}$                 & \\
      & $\ftail$               & $0.083 ^{+0.083}_{-0.045}$              & \\
      & $\sigol$ (ps)          & $36.2 ^{+5.0}_{-3.5}$                   & \\
      & $\fol^{\rm multiple}$ & $(5.83 ^{+3.02}_{-2.25})\times 10^{-4}$ & \\
      & $\fol^{\rm single}$   & $0.0306 \pm 0.0036$                     &\\
\hline
\hline
    \end{tabular}
\end{center}
  \end{table}

\begin{figure}
\begin{center}
  \resizebox{0.6\textwidth}{!}{\includegraphics{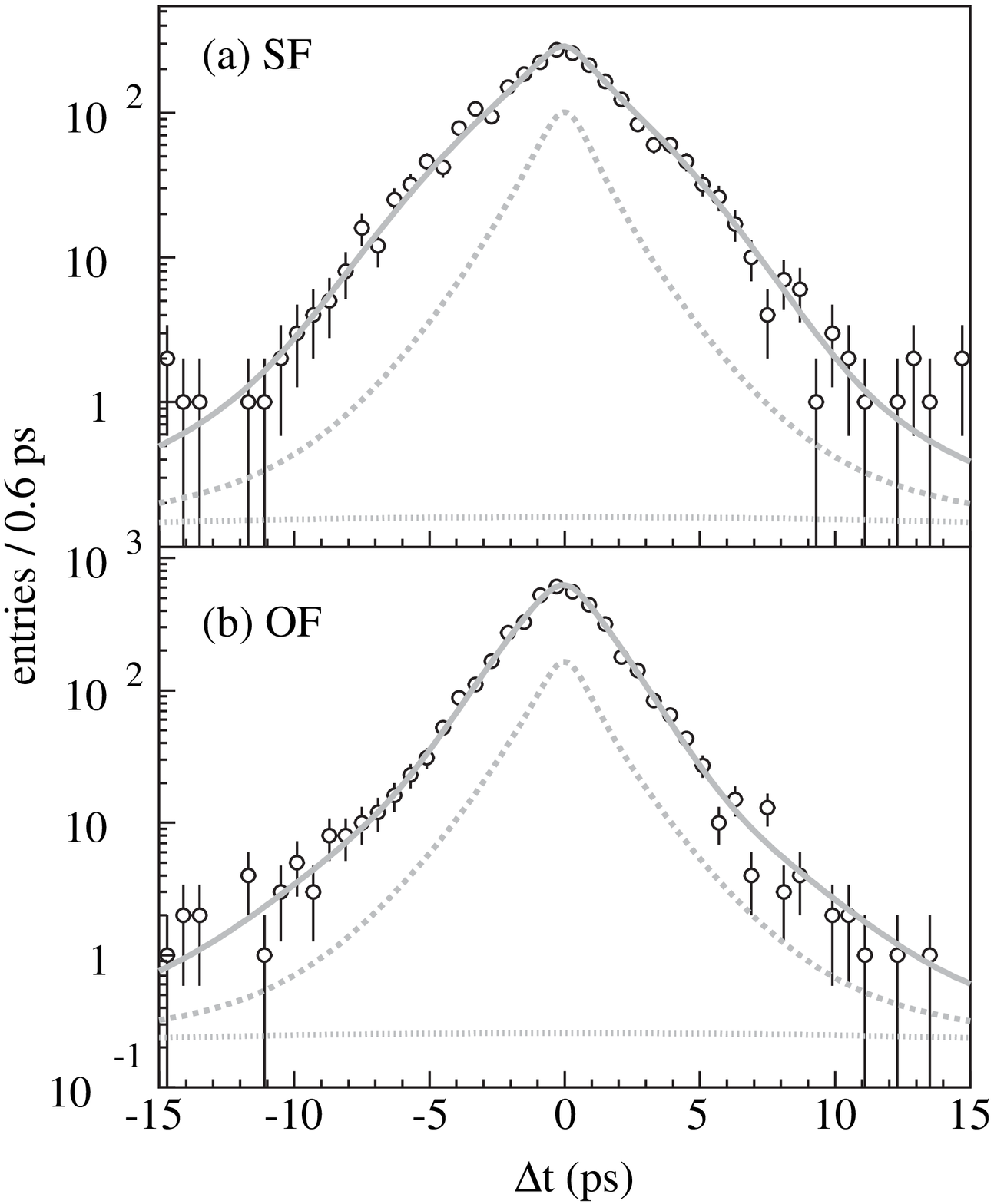}}%
\end{center}
  \caption{\label{fig:mixing}
    The $\dt$ distributions for (a) SF and (b) OF events with
the fitted curves superimposed.
The dashed, dotted, and solid curves show the background, outliers, and the sum of
backgrounds and signal, respectively.}
\end{figure}

The same data sample is used to determine the $B^0\overline{B}{}^0$ oscillation frequency
$\Delta m_d$ from the time evolution of opposite-flavor (OF; $B^0\overline{B}{}^0$)
and same-flavor (SF; $B^0B^0$, $\overline{B}{}^0\overline{B}{}^0$) neutral $B$
decays~\cite{mixing}.  The signal PDF is defined as the convolution of the true PDF,
\begin{eqnarray}
\mathcal{P}^{\rm OF}(\Delta t) &= & \frac{1}{4\tau_{\overline{B}{}^0}}\exp\left(-\frac{|\Delta t|}{\tau_{\overline{B}{}^ 0}}\right)[1+(1-2w)\cos(\Delta {m_d}\Delta t)]\ \  {\rm and} \nonumber\\
\mathcal{P}^{\rm SF}(\Delta t) &= & \frac{1}{4\tau_{\overline{B}{}^0}}\exp\left(-\frac{|\Delta t|}{\tau_{\overline{B}{}^ 0}}\right)[1-(1-2w)\cos(\Delta {m_d}\Delta t)], 
\end{eqnarray}
with the resolution function $\Rsig(\Delta t)$.
Here $w$ is the probability for an incorrect flavor assignment and determined simultaneously with
$\Delta m_d$ by the fit.
Figure~\ref{fig:mixing} shows the $\Delta t$ distributions for OF and SF events with fitted curves 
superimposed.  
The fit yields
$$\Delta m_d = 0.528\pm 0.017({\rm stat})\pm 0.011({\rm syst})\ \ {\rm ps}^{-1}.$$
The error arising from the resolution function accounts for $0.009$~ps$^{-1}$ of the total systematic error quoted above.

\subsection{Time-dependent $CP$ asymmetry}

The resolution function obtained above is also used for  
the measurement of mixing-induced  $CP$  violation in the neutral $B$ meson system.
The Standard Model predicts a $CP$-violating asymmetry in the time-dependent
rates for $B^0$ and $\overline{B}{}^0$ decays to a common $CP$ eigenstate $f_{CP}$,
where the transition is dominated by the $b\to c\bar{c}s$ process:
\begin{equation}
  A(t) \equiv \frac{\Gamma(\bzb \to \fcp) - \Gamma(\bz \to \fcp)}
  {\Gamma(\bzb \to \fcp) + \Gamma(\bz \to \fcp)}
  = -\xi_f \sin2\phi_1\sin(\Delta m_d t),
\end{equation}
where $\Gamma(\bz,\bzb \to \fcp)$ is the decay rate for $\bz$ or $\bzb$
to $\fcp$ at a proper time $t$ after production,
$\xi_f$ is the $CP$ eigenvalue of $\fcp$,
and $\sin2\phi_1$ is the $CP$ violation parameter.
A non-zero value for $\sin2\phi_1$ establishes that $CP$ symmetry is violated 
in the neutral $B$ meson system.
We use events in which  one of the $B$ mesons decays to $\fcp$ at time $\tcp$,
and the other decays to a self-tagging state $\ftag$,
which distinguishes $\bz$ from $\bzb$, at time $\ttag$.
The $CP$ violation manifests itself as an asymmetry $A(\Dt)$,
where $\Dt$ is the proper time interval
between the two decays: $\Dt \equiv \tcp - \ttag$.

The PDF expected for the signal distribution is
\begin{equation}
  \label{eq:deltat}
  \mathcal{P}_{\rm sig}(\Dt, q, w, \xi_f) =
  \frac{1}{4\taubz}\exp\left(-\frac{|\Dt|}{\taubz}\right)[1 - q\xi_f(1-2w)\sinbb\sin(\dM\Dt)],
\end{equation}
where $q$ has the discrete value $ +1 (-1)$ when the tag-side $B$ meson is
likely to be a $B^0$ ($\overline{B}{}^0$), and $w$ is, as mentioned in the previous section,
the probability for
an incorrect flavor assignment (wrong-tag probability)~\cite{tagging}.
The PDF is convolved with $\Rsig(\Dt)$
to determine the likelihood value for each event as a function of $\sinbb$:
\begin{eqnarray}
  P_i &=& (1-\fol)[ \fsig \int  \mathcal{P}_{\ sig}(\Dt',q,w,\xi_f)\Rsig(\Dt-\Dt'){\rm d}\Dt'
  \nonumber \\
  && \quad + \; (1-\fsig)P_{\rm bkg}(\Dt)] 
  + \fol \Pol(\Dt).
\end{eqnarray}
The only free parameter in the final fit is $\sinbb$.
The quality of the vertex fit for the tag-side $B$ meson (or associated $B$ meson) can be
correlated with $w$, and, therefore, the resolution function  can also be correlated with $w$ primarily because of 
smearing due to non-primary tracks.
$\Rnp$ is designed to account for this correlation by making  the parameters of vertex position shifts ($\tau_{\rm np}^{\rm p}$ and  $\tau_{\rm np}^{\rm n}$) 
a function of vertex-fit qualities ($\sigma_{\rm acs}$ and $\xi_{\rm asc}$).
MC simulation study shows that the remaining $w$ dependence is negligible.

\begin{figure}
\begin{center}
  \resizebox{0.6\textwidth}{!}{\includegraphics{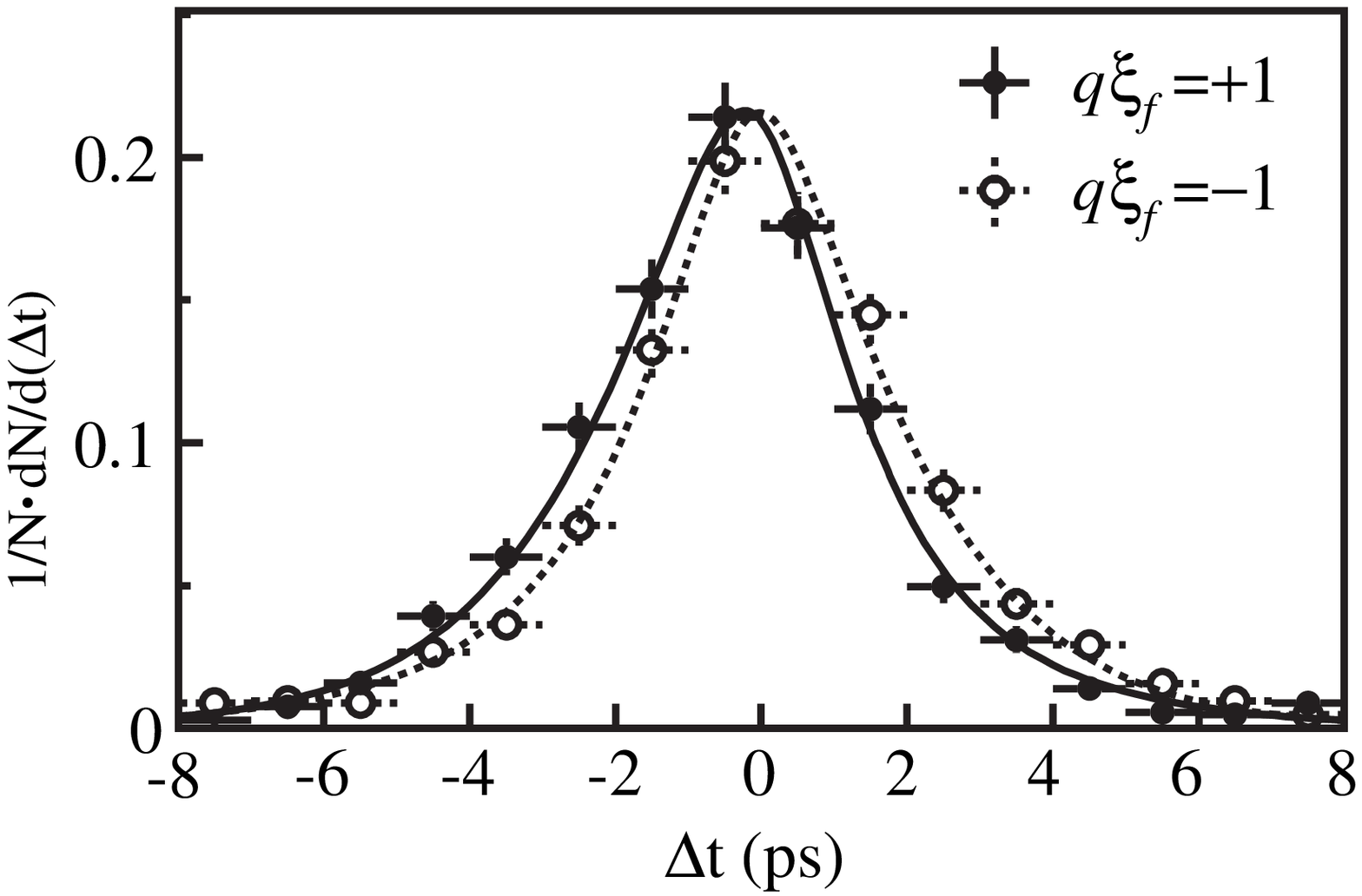}}%
\end{center}
  \caption{\label{fig:cpfit}
    The $\dt$ distributions for the events with $q\xi_f=+1$ (solid points) and
$q\xi_f=-1$ (open points). The results of the global fit with $\sin2\phi_1=0.719$ are
shown as solid and dashed curves, respectively.
See text for the detail.}
\end{figure}

The value of $\sin 2\phi_1$ is measured using a $78~{\rm fb}^{-1}$ data sample, which contains 
85 million $B\overline {B}$ pairs~\cite{CPV}.
This data sample has been analyzed using a new track reconstruction algorithm that
provides better performance in vertex reconstruction.
We repeat the lifetime fit to this sample to determine the parameter values for $\Rsig(\Delta t)$.
We find the fraction $f_{\rm tail}$ for the tail part of the detector resolution  is consistent  with zero,
and therefore set $f_{\rm tail}=0$ for this data sample\footnote{Consequently, $s_{\rm tail}$ is not used}. 
In addition,  the improvement in statistics enables us to determine some parameters that are previously
determined only by MC simulations.
The fractions of the prompt component in $\Rnp$ for multiple- and single-track vertices
($f_{\delta,\overline{B}{}^0}^{\rm multiple}$,  $f_{\delta,\overline{B}{}^0}^{\rm single}$ for
$\overline{B}{}^0$ mesons, and $f_{\delta,B^-}^{\rm multiple}$,  $f_{\delta,B^-}^{\rm single}$ for
$B^-$ mesons)
are  determined by the lifetime fit to the data.
The values so obtained are consistent with the values determined using MC simulations.
Table~\ref{tab:newparams} lists the parameter values  for $\Rsig(\Delta t)$.
We find the resulting $\Delta t$ resolution to be  $\sim 1.43~{\rm ps}$ (rms),
improved over the resolution  of $\sim 1.56~{\rm ps}$  obtained for the $29~{\rm fb}^{-1}$ sample.
Figure~\ref{fig:cpfit} shows the observed $\Delta t$ distributions for $q\xi_f=+1$ (solid points) and
$q\xi_f=-1$ (open points) event samples. 
The asymmetry between the two  distributions is proportional to $\sin 2\phi_1$ and
demonstrates the violation of $CP$ symmetry. 
The value of $\sin 2\phi_1$ is measured to be
%\begin{equation}
$$\sin 2\phi_1 = 0.719 \pm 0.074 {\rm (stat)} \pm 0.035 {\rm (syst)}.$$
The quoted systematic error includes  
the systematic error due to the uncertainty in the resolution function ($0.014$). 
%\end{equation}

 \begin{table}
\begin{center}
\caption{\label{tab:newparams}The resolution function parameters determined
using a $78~{\rm fb}^{-1}$ data sample.
This data sample has been analyzed using a new track reconstruction algorithm.}  
    \begin{tabular}{clcc}
\hline
\hline
      & Parameters             & Values                                  & \\
      \hline
      & $\taubz$ (ps)          & $1.551 \pm 0.018$                       & \\
      & $\taubm$ (ps)          & $1.658 \pm 0.016$                       & \\
      \hline
      & $\srec^0$              & $0.987 ^{+0.117}_{-0.124}$              & \\
      & $\srec^1$              & $0.094 \pm 0.008$                       & \\
      & $\sasc^0$              & $0.778 \pm 0.048$                       & \\
      & $\sasc^1$              & $0.044 \pm 0.002$                       & \\
      & $\smain$               & $0.972 \pm 0.045$                       & \\
      & $\sigol$ (ps)          & $42.0 ^{+4.6}_{-3.5}$                   & \\
      & $\fol^\text{multiple}$ & $(1.65 ^{+1.13}_{-0.82})\times 10^{-4}$ & \\
      & $\fol^\text{single}$   & $0.0269 \pm 0.0019$           & \\
      & $f_{\delta,\bzb}^\text{multiple}$ & $0.555 \pm 0.042$   & \\
      & $f_{\delta,\bm}^\text{multiple}$  & $0.440 \pm 0.046$   & \\
      & $f_{\delta,\bzb}^\text{single}$   & $0.701 \pm 0.040$   & \\
      & $f_{\delta,\bm}^\text{single}$    & $0.764 \pm 0.044$   &\\
\hline
\hline
    \end{tabular}
\end{center}
 \end{table}

%% file: conclusion.tex
\section{Conclusions}
\label{sec:conclusion}
The resolution function, which is used in an unbinned maximum likelihood fit for the time-dependent measurements at the Belle experiment, 
is studied in detail.
The resolution function  is described as a convolution of three components;
the detector resolution,
the smearing due to non-primary tracks, and
the kinematic approximation.
The functional forms to describe these components are determined based on 
detailed MC simulation studies.
The parameters for the detector resolution are determined using the data.
The resulting  resolution function has successfully described the $\Delta t$ distribution 
used for measurements
of $B$ meson lifetimes,
$\bz\bzb$ oscillation frequency, and
the mixing-induced \CP\ asymmetry parameter $\sin2\phi_1$.

%% file: biblio.tex
\newcommand*{\prl}[3]{Phys.\ Rev.\ Lett.\ \textbf{#1}, #2 (#3)}
\newcommand*{\prd}[3]{Phys.\ Rev.\ D \textbf{#1}, #2 (#3)}
\newcommand*{\plb}[3]{Phys.\ Lett.\ B \textbf{#1}, #2 (#3)}